\documentstyle{article}
\begin{document}

\begin{titlepage}

\title{New relations in the algebra of the Baxter Q-operators.}
\author{A. A. Belavin, A. V. Odesskii, R. A. Usmanov \\ {} \\
              {\it Landau Institute} \\
              {\it for Theoretical Physics}\\
              {\it Chernogolovka, Moscow region, 142432,}\\
              {\it Russia}}
\date{hep-th/0110126}

\end{titlepage}

\maketitle

\begin{abstract}
We consider irreducible cyclic representations of the algebra of
monodromy matrices corresponding to the $R$-matrix of the
six-vertex model. In roots of unity the Baxter Q-operator can be
represented as a trace of a tensor product of $L$-operators
corresponding to one of these cyclic representations and
satisfies the $TQ$-equation. We find a new algebraic structure
generated by these $L$-operators, and as a consequence, by the
$Q$-operators.
\end{abstract}

\section*{Introduction.}

In his papers \cite{kn:gnusQ1} Baxter introduced the $Q$-operator
and used it for solving of the six-vertex model. The
$Q$-operators form a family that commutes with a family of the
transfer-matrices $T(u)$, with the $TQ$-equation satisfying. The
latter equation relates two families with each other and is a key
for solving the model.

In \cite{kn:gnus1} (see also \cite{kn:gnus4}, \cite{kn:gnus5})
the expressions for the Boltzmann weights of the chiral Potts
model were found, that are solutions of the star-triangle
relation. The $R$-matrix $S$ of the model can be represented as a
product of four such Boltzmann weights.

Light on the algebraic structure of the $Q$-operators in the
particular case of the six-vertex model and its relation to the
$R$-matrix of the chiral Potts model shed the paper by Bazhanov
and Stroganov \cite{kn:gnus2}. In the $N$-th roots of unity ($N$
is a prime number) they found the $N$-dimensional cyclic
representation $\cal L$ of the Yang--Baxter algebra related to the
$R$-matrix of the usual six-vertex model. The trace over the
$N$-dimensional quantum space of a tensor product of
$L$-operators possesses the properties of the $Q$-operator. In
particular, it satisfies the $TQ$-equation.

In the paper by Tarasov \cite{kn:gnus3} irreducible cyclic
representations of the algebra of monodromy matrices
corresponding to the $R$-matrix of the six-vertex model in roots
of unity were described.

Recently the $Q$-operator came to the centre of attention again.
For some models of statistical physics it was shown
\cite{kn:Backlund1}, \cite{kn:Backlund2}, \cite{kn:Sklyanin},
\cite{kn:Pronko} that the $Q$-operator is a quantum analogue of
the B$\ddot{a}$cklund transformation.

In the paper \cite{kn:gnus2} mentioned above the $R$-matrix of the
chiral Potts model $S$ was derived as an operator intertwining
tensor products of two cyclic representations of the algebra of
monodromy matrices, with they multiplying at first in one order
and then in the inverse order.

As it was shown by A.Odesskii (unpublished), the four factors
generating the $R$-matrix of the chiral Potts model $S$ are
actually intertwiners that provide some elementary isomorphisms
of cyclic representations of the $L$-operators algebra.

In this work we clarify the conditions under which two cyclic
representations are equivalent and find the corresponding
intertwiner. We also solve the same problem for two tensor
products of a pair of cyclic representations. The obtained
intertwiners are a generalization of the well-known vertex
weights of the chiral Potts model and satisfy some modification
of the star-triangle equations.

The plan of the paper is the following. In Section 1 we introduce
the notion of cyclic representations of the algebra of
$L$-operators. In Section 2 we discuss different definitions of
cyclic representations. In Section 3 we derive the $TQ$-equation.
In Section 4 we discuss some special cases of elementary
isomorphisms acting on cyclic representations of the algebra of
monodromy matrices. In Section 5 we find them in the general case.
We show that the intertwiners of these elementary isomorphisms
satisfy a generalized star-triangle relation of the chiral Potts
model. In Section 6 we write some relations in the algebra of
$Q$-operators. In Section 7 we discuss some questions related to
possible future investigations. In Appendices we prove some
formulae used in the paper.

\section{Cyclic representations of the Yang--Baxter algebra.}

Following to \cite{kn:gnus3}, we introduce the definition of the
$R$-matrix:
\begin{equation}
R(u) = \left[
\begin{array}{llll}
1- u \omega & & & \\
\vspace{-2mm}\\
& \omega(1-u) & u(1-\omega) & \\
\vspace{-2mm}\\
& 1-\omega & 1-u & \\
& & & 1-u \omega
\end{array}
\right].
\label{eq:111}
\end{equation}
It is connected with the algebra $U_{q}(sl_{2})$ \cite{kn:gnus6},
\cite{kn:gnus7} and can be obtained from the $R$-matrix of the
usual six-vertex model by means of a simple tranformation (see
Section 2). For short we denote ${\cal M}=End \; {\bf C}^{2}$ so
that $R(u) = {\cal M} \otimes {\cal M}$.

The algebra of monodromy matrices $\cal A$ is the algebra with
generators $A(u)$, $B(u)$, $C(u)$, $D(u)$, $H$, $H^{-1}$ and
relations
\begin{equation}
\begin{array}{l}
R(u) \stackrel{1}{L}(u v) \stackrel{2}{L}(v) = \stackrel{2}{L}(v)
\stackrel{1}{L}(u v) R(u),\\
\vspace{-2mm}\\
\left[ \hat{\omega} \otimes H, L(u) \right] =0, \qquad H H^{-1}=H^{-1}H=1,\\
\vspace{-2mm}\\
L(u) = \left[
\begin{array}{ll}
A(u) & B(u)\\
\vspace{-2mm}\\
C(u) & D(u)
\end{array}
\right] \in {\cal M} \otimes {\cal A}, \qquad \hat{\omega} =
diag(1, \omega).
\end{array}
\label{eq:112}
\end{equation}
The indices '1' and '2' over $L$ denote a two-dimensional space in
which the corresponding $L$-operator multiplies by the $R$-matrix.
At that both $L$-operators act in the same auxiliary space.

As indicated in \cite{kn:gnus3}, in the algebra $\cal A$ one can
introduce the coproduct $\Delta$:
$$
\begin{array}{l}
\Delta(L(u)) = L_{1}(u) L_{2}(u) \in {\cal M} \otimes {\cal A}
\otimes {\cal A},\\
\vspace{-2mm}\\
\Delta(H) = H \otimes H.
\end{array}
$$
By lower indices '1' and '2' are denoted the quantum spaces, in
which the corresponding $L$-operators act. At that the
$L$-operators considering as two-dimensional matrices (each
matrix element is an operator in one of the two quantum spaces),
multiply with each other according to the usual rule of
multiplication of matrices.

Thus a tensor product of some representations of the algebra
$\cal A$ is a representation of $\cal A$ too.

Let us now define the quantum determinant:
\begin{equation}
{\det}_{q} \; L(u) = D(u)A(u \omega^{-1}) - C(u)B(u \omega^{-1}).
\label{eq:113}
\end{equation}
One can verify that $H^{-1} {\det}_{q} \; L(u)$ is a central
element of the algebra $\cal A$.

Hereafter we set $\omega^{N}=1$. As shown in \cite{kn:gnus3} in
this case the center of the algebra $\cal A$ increases, namely,
the following operators become central:
$$
\langle {\cal O} \rangle (u) = \prod_{k=0}^{N-1} {\cal O}(u
\omega^{k}), \qquad {\cal O} = A,B,C,D,
$$
and thus we can define the matrix of central elements:
$$
\langle L \rangle = \left[
\begin{array}{ll}
\langle A \rangle & \langle B \rangle\\
\vspace{-2mm}\\
\langle C \rangle & \langle D \rangle
\end{array}
\right].
$$

One can show \cite{kn:gnus3} that $L = L_{1}L_{2}$ satisfies the
equation:
$$
\Delta(\langle L \rangle) = \langle L_{1} \rangle \langle L_{2}
\rangle, \qquad \langle {\det}_{q} \; L  \rangle = \det \langle L
\rangle.
$$

The $N$-dimensional cyclic representation $\pi$ of the algebra
$\cal A$ has the form \cite{kn:gnus3}:
\begin{equation}
\begin{array}{l}
L(u,p_{1},p_{2}) = \left(
\begin{array}{ll}
c_{1} c_{2} Z - b_{1} b_{2} u & -u(b_{1}d_{2} - c_{1}a_{2}Z)X\\
\vspace{-2mm}\\
X^{-1}(d_{1}b_{2}-a_{1}c_{2}Z) & d_{1}d_{2}-a_{1}a_{2}\omega u Z
\end{array}
\right)\\
\vspace{-2mm}\\
H_{\pi} = h Z, \qquad p_{i}=(a_{i},b_{i},c_{i},d_{i}), i=1,2.
\end{array}
\label{eq:114}
\end{equation}

The action of the operators $X$, $Z$ on the standard basis in
${\bf C}^{N}$ reads as follows:
$$
Z |k \rangle = \omega^{k}|k\rangle, \qquad X|k\rangle =
|k+1\rangle, \qquad (k=0,...,N-1, |N\rangle \equiv |0\rangle).
$$

We also have
$$
\langle L(p_{1},p_{2}) \rangle (v) =
\left(
\begin{array}{ll}
c_{1}^{N} c_{2}^{N} - b_{1}^{N} b_{2}^{N} v &
-v(b_{1}^{N}d_{2}^{N} - c_{1}^{N}a_{2}^{N})\\
\vspace{-2mm}\\
d_{1}^{N}b_{2}^{N} - a_{1}^{N}c_{2}^{N} & d_{1}^{N}d_{2}^{N} -
a_{1}^{N}a_{2}^{N}v
\end{array}
\right). $$ Though in formulae (\ref{eq:114}) there are eight (in
addition to $v$) parameters, the $N$-dimensional representation
depends only on six of them:\\
the substitution
$$ a_{1} \rightarrow \lambda a_{1}, \qquad a_{2}
\rightarrow \lambda^{-1} a_{2},
$$
$$
c_{1} \rightarrow \lambda c_{1}, \qquad c_{2} \rightarrow
\lambda^{-1} c_{2},
$$
$$
b_{1} \rightarrow b_{1}, \qquad b_{2} \rightarrow b_{2}, \qquad
d_{1} \rightarrow d_{1}, \qquad d_{2} \rightarrow d_{2},
$$
where $\lambda$ is an arbitrary number, does not change the
operator $L(u,p_{1},p_{2})$. The same is true for the substitution
$$
b_{1} \rightarrow \lambda b_{1}, \qquad b_{2} \rightarrow
\lambda^{-1} b_{2},
$$
$$
d_{1} \rightarrow \lambda d_{1}, \qquad d_{2} \rightarrow
\lambda^{-1} d_{2},
$$
$$
a_{1} \rightarrow a_{1}, \qquad a_{2} \rightarrow a_{2}, \qquad
c_{1} \rightarrow c_{1}, \qquad c_{2} \rightarrow c_{2}.
$$

Apart from, the projective equivalence class of the $L$-operators
depends only on four additional parameters, since
$$
L(\lambda
p_{1}, p_{2}) = \lambda L(p_{1}, p_{2}), \qquad L(p_{1}, \mu
p_{2}) = \mu L(p_{1}, p_{2}),
$$
where $\lambda$, $\mu$ are
arbitrary numbers.

Two representations $L_{1}(u,p_{1},p_{2}) L_{2}(u,p_{3},p_{4})$
and $L_{1}(u,p_{3},p_{4}) L_{2}(u,p_{1},p_{2})$ are equivalent if
and only if one can choose $p_{i}$, $i=1,2,3,4$, satisfying the
conditions
\begin{equation}
\frac{a_{i}^{N} \pm b_{i}^{N}}{c_{i}^{N} \pm d_{i}^{N}} =
\lambda_{\pm},
\label{eq:ququ}
\end{equation}
where $\lambda_{\pm}$ independent of $i$ (Appendix \ref{Ferma}).
The intertwiner given by the equation
\begin{equation}
\begin{array}{l}
{\bf S}(p_{1},p_{2},p_{3},p_{4})L_{1}(u,p_{1},p_{2})
L_{2}(u,p_{3},p_{4})=\\
\vspace{-2mm}\\
=L_{1}(u,p_{3},p_{4}) L_{2}(u,p_{1},p_{2}) {\bf
S}(p_{1},p_{2},p_{3},p_{4}),
\end{array}
\label{eq:115}
\end{equation}
has explicit expression through Boltzmann weights of the chiral
Potts model $W_{p q}$, $\overline{W}_{p q}$:
\begin{equation}
\begin{array}{l}
{\bf S}(p_{1},p_{2},p_{3},p_{4}) = F(p_{1},p_{4};
X_{1}X_{2}^{-1}) G(p_{1},p_{3}; Z_{1})\\
\vspace{-2mm}\\
\qquad \qquad \qquad \qquad \times G(p_{2},p_{4}; Z_{2})
F(p_{2},p_{3};
X_{1}X_{2}^{-1}),\\
\vspace{-2mm}\\
\vspace{-2mm}\\
\displaystyle G(p,q; \omega^{k}) = W_{p q}(k), \qquad F(p,q;
\omega^{k}) = \sum_{l=1}^{N} \omega^{k l} \overline{W}_{p q}(l).
\end{array}
\label{eq:116}
\end{equation}

\section{Cyclic representations of the Yang--Baxter algebra in the
form of Bazhanov--Stroganov.}

Parallel with the $L$-operators introduced in the previous section
one can consider their version related to another choice of the
$R$-matrix.

Let us consider the usual $R$-matrix of the ice model,
\begin{equation}
R_{ice}(x)= \left[
\begin{array}{llll}
x \omega_{1} - x^{-1} \omega_{1}^{-1} & & &\\
\vspace{-2mm}\\
& x - x^{-1} & \omega_{1} - \omega_{1}^{-1} &\\
\vspace{-2mm}\\
& \omega_{1} - \omega_{1}^{-1} & x - x^{-1} &\\
\vspace{-2mm}\\
& & & x \omega_{1} - x^{-1} \omega_{1}^{-1}
\end{array}
\right],
\label{eq:b2}
\end{equation}
and corresponding relations in the Yang--Baxter algebra,
\begin{equation}
R_{ice}(x) \stackrel{1}{L}_{ice}(x y) \stackrel{2}{L}_{ice}(y) =
\stackrel{2}{L}_{ice}(y) \stackrel{1}{L}_{ice}(x y) R_{ice}(x).
\label{eq:a2}
\end{equation}

The $N$-dimensional representation of the algebra (\ref{eq:a2})
one can write in the following form:
$$
L_{ice}(y,p_{1},p_{2}) = \left[
\begin{array}{ll}
y^{-1}c_{1}c_{2}Z_{1} - b_{1}b_{2}y Z_{1}^{-1} &
-(b_{1}d_{2}Z_{1}^{-1}-c_{1}a_{2}Z_{1})X\\
\vspace{-2mm}\\
\omega_{1} X^{-1}(d_{1}b_{2}Z_{1}^{-1}-a_{1}c_{2}Z_{1}) &
y^{-1}d_{1}d_{2}Z_{1}^{-1}-a_{1}a_{2}\omega_{1}^{2} y Z_{1}
\end{array}
\right], \label{eq:c2}
$$
where the operators $Z_{1}$, $X$ satisfy
the relations
$$
Z_{1}^{N} = 1, \qquad X^{N} = 1, \qquad Z_{1}X =
\omega_{1}X Z_{1}.
$$

Let us make the following substitution: $$ R_{1 2} \rightarrow
C_{1}^{-1}(x y)C_{2}^{-1}(y)R_{1 2}C_{2}(y) C_{1}(x y), \qquad
L_{1}(x y) \rightarrow C_{1}(x y)^{-1} L_{1} (x y) C_{1}(x y), $$
$$ L_{2}(y) \rightarrow C_{2}^{-1}(y) L_{2}(y) C_{2}(y), $$ where
$$ C(y)= \left[
\begin{array}{ll}
1 & 0\\
\vspace{-2mm}\\
0 & y
\end{array}
\right], $$ and the index of the matrix $C$ indicates the space
this matrix acts in.

As a result of this substitution, the equation $R L L = L L R$ is
valid, as before. At that
\begin{equation}
{\cal R}(x)= \left[
\begin{array}{llll}
x \omega_{1} - x^{-1} \omega_{1}^{-1} & & &\\
\vspace{-2mm}\\
& x - x^{-1} & x(\omega_{1} - \omega_{1}^{-1}) &\\
\vspace{-2mm}\\
& x^{-1}(\omega_{1} - \omega_{1}^{-1}) & x - x^{-1} &\\
\vspace{-2mm}\\
& & & x \omega_{1} - x^{-1} \omega_{1}^{-1}
\end{array}
\right],
\label{eq:d2}
\end{equation}
and the $L$-operator is given by the formula:
\begin{equation}
{\cal L}(y,p_{1},p_{2}) = \left[
\begin{array}{ll}
y^{-1}c_{1}c_{2}Z_{1}-b_{1}b_{2}y Z_{1}^{-1} &
-y(b_{1}d_{2}Z_{1}^{-1}-c_{1}a_{2}Z_{1})X\\
\vspace{-2mm}\\
\omega_{1} y^{-1} X^{-1}(d_{1}b_{2}Z_{1}^{-1}-a_{1}c_{2}Z_{1}) &
y^{-1}d_{1}d_{2}Z_{1}^{-1}-a_{1}a_{2}\omega_{1}^{2} y Z_{1}
\end{array}
\right].
\label{eq:e2}
\end{equation}

It is the last $L$-operator that was found in the paper
\cite{kn:gnus2}. We call it the cyclic representation of the
algebra of monodromy matrices in the form of Bazhanov--Stroganov.

Let us now miltiply $L(y)$ by $y Z_{1}$ and introduce the
following notations: $$ v = y^{2}, \qquad Z = Z_{1}^{2}, \qquad
\omega = \omega_{1}^{2}. $$ Then we obtain the operator
(\ref{eq:114}). At that the $R$-matrix becomes (in order the
equation $R L L = L L R$ being valid as before): $$ R(u) = -
\omega_{1} x K_{1} R_{BS}(x) K_{2}^{-1}, $$ where $u=x^{2}$, $$ K
= \omega_{1}^{(\sigma^{z} - 1)/2} = \left(
\begin{array}{ll}
1 & 0\\
\vspace{-2mm}\\
0 & \omega_{1}^{-1}
\end{array}
\right). $$ The lower index of the matrix $K$ is denoted the
space it acts in. As it can be seen without difficult the matrix
$R$ coincides with (\ref{eq:111}) and the operator $L$ coincides
with (\ref{eq:114}).

We emphasize that hereinafter we denote the cyclic
representations in the form of Tarasov by $L$ and the cyclic
representations in the form of Bazhanov--Stroganov by $\cal L$ in
order to avoid a confusion.

Matrix elements of the $L$-operators we denote by $L_{i
\alpha}^{j \beta}$ and ${\cal L}_{i \alpha}^{j \beta}$, $i,j =
0,1$, $\alpha, \beta = 0,...,N-1$, correspondingly.

\section{The $Q$-operator and the $TQ$-equation.}

The transfer matrix built with the aim of ${\cal L}(u)$, $$ {\cal
Q}(u) = {\rm tr}_{0} {\cal L}_{1 0}(u){\cal L}_{2 0}(u)...{\cal
L}_{n 0}(u), $$ where the trace is calculated in the
$N$-dimensional space, possesses a very important property which
allows to say about it as about the Baxter $Q$-operator
\cite{kn:gnus2}. Namely, it satisfies the $TQ$-equation. Let us
prove this statement.

We consider the equation
\begin{equation}
{\cal R}_{i_{1} i_{2}}^{j_{1} j_{2}}(u) {\cal L}_{j_{1}
\alpha}^{k_{1} \beta}(u v) {\cal L}_{j_{2} \beta}^{k_{2}
\gamma}(v) = {\cal L}_{i_{2} \alpha}^{j_{2} \beta}(v) {\cal
L}_{i_{1} \beta}^{j_{1} \gamma}(u v) {\cal R}_{j_{1} j_{2}}^{k_{1}
k_{2}}(u). \label{eq:new}
\end{equation}
It can be shown graphically as in Fig.1.

\begin{figure}
\begin{center}
\begin{picture}(220,220)(0,0)

\put(40,163){\line(1,0){40}} \put(60,143){\line(0,1){40}}

\put(40,163){\vector(1,0){13}} \put(40,163){\vector(1,0){33}}
\put(60,143){\vector(0,1){13}} \put(60,143){\vector(0,1){33}}

\put(165,163){\line(1,0){40}} \put(185,143){\line(0,1){40}}
\put(187,143){\line(0,1){40}}

\put(165,163){\vector(1,0){13}} \put(165,163){\vector(1,0){33}}
\put(185,143){\vector(0,1){13}} \put(187,143){\vector(0,1){13}}
\put(185,143){\vector(0,1){33}} \put(187,143){\vector(0,1){33}}

\put(51,153){$u$} \put(176,153){$v$}

\put(-12,160){$R_{i_{1} i_{2}}^{j_{1} j_{2}}(u) \, =$}
\put(120,160){${\cal L}_{i \alpha}^{j \beta}(v) \, =$}

\put(41,166){$i_{1}$} \put(73,154){$j_{1}$} \put(62,144){$i_{2}$}
\put(63,174){$j_{2}$}

\put(167,166){$i$} \put(200,154){$j$} \put(190,145){$\alpha$}
\put(190,174){$\beta$}

\put(60,163){\oval(8,8)[bl]} \put(185,163){\oval(8,8)[bl]}

\put(150,20){\line(0,1){90}} \put(152,20){\line(0,1){90}}
\put(130,30){\line(2,1){90}} \put(130,100){\line(2,-1){90}}

\put(130,30){\vector(2,1){13}} \put(130,30){\vector(2,1){45}}
\put(130,30){\vector(2,1){83}} \put(130,100){\vector(2,-1){13}}
\put(130,100){\vector(2,-1){45}} \put(130,100){\vector(2,-1){83}}

\put(150,20){\vector(0,1){13}} \put(150,20){\vector(0,1){45}}
\put(150,20){\vector(0,1){83}} \put(152,20){\vector(0,1){13}}
\put(152,20){\vector(0,1){45}} \put(152,20){\vector(0,1){83}}

\put(105,62){$=$}

\put(70,20){\line(0,1){90}} \put(68,20){\line(0,1){90}}
\put(0,53){\line(2,1){90}} \put(0,75){\line(2,-1){90}}

\put(0,53){\vector(2,1){13}} \put(0,53){\vector(2,1){45}}
\put(0,53){\vector(2,1){83}} \put(0,75){\vector(2,-1){13}}
\put(0,75){\vector(2,-1){45}} \put(0,75){\vector(2,-1){83}}
\put(70,20){\vector(0,1){13}} \put(70,20){\vector(0,1){45}}
\put(70,20){\vector(0,1){83}} \put(68,20){\vector(0,1){13}}
\put(68,20){\vector(0,1){45}} \put(68,20){\vector(0,1){83}}

\put(148,10){$3$} \put(122,27){$2$} \put(122,98){$1$}
\put(140,23){$v$} \put(134,80){$u v$} \put(184,63){$u$}
\put(154,25){$\alpha$} \put(154,63){$\beta$}
\put(154,100){$\gamma$} \put(134,40){$i_{2}$}
\put(175,45){$j_{2}$} \put(222,75){$k_{2}$} \put(134,105){$i_{1}$}
\put(175,85){$j_{1}$} \put(222,50){$k_{1}$}

\put(58,10){$3$} \put(-8,50){$2$} \put(-8,73){$1$}
\put(57,73){$v$} \put(53,32){$u v$} \put(3,63){$u$}
\put(73,23){$\alpha$} \put(74,63){$\beta$} \put(74,103){$\gamma$}
\put(5,48){$i_{2}$} \put(40,80){$j_{2}$} \put(82,86){$k_{2}$}
\put(5,80){$i_{1}$} \put(40,42){$j_{1}$} \put(82,37){$k_{1}$}

\end{picture}
\end{center}
\caption{}
\end{figure}

If the indices $i_{1}$, $k_{1}$ are fixed, the equation
(\ref{eq:new}) is an operator in the tensor product ${\bf C}^{2}
\times {\bf C}^{N}$. Let us act by this operator on the vector
$\psi_{k_{2} \gamma}$, that belongs to the kernel of the operator
${\cal L}_{2 3}(v)$, that is, on the vector satisfying the
equation $$ {\cal L}_{j_{2} \beta}^{k_{2} \gamma}(v) \psi_{k_{2}
\gamma} = 0. $$ A nontrivial kernel of ${\cal L}_{2 3}(v)$ there
exists only if the spectral parameter $v$ equals some special
values $v=v_{*}$: $$ v_{*}^{2} = \frac{c_{1}d_{1}}{a_{1}b_{1}} $$
or $$ v_{*}^{2} = \frac{c_{2}d_{2}}{a_{2}b_{2}}. $$

One can see from Fig.1 that the kernel of the operator ${\cal
L}_{2 3}(v_{*})$ is a subspace which is invariant with regard to
the tensor product ${\cal L}_{1 3}(u v_{*}) {\cal R}_{1 2}(u)$.
In this case the supplement of the kernel considering as a coset
space is an invariant space too. Therefore the matrix of the
operator ${\cal L}_{1 3}(u v_{*}) {\cal R}_{1 2}(u)$ has a
block-diagonal form: $$ {\cal L}_{1 3}(u v_{*}) {\cal R}_{1 2}(u)
= \left[
\begin{array}{ll}
P_{1} & *\\
\vspace{-2mm}\\
0 & P_{2}
\end{array}
\right], $$ where all blocks are $N$-dimensional matrices, and  we
denote the matrix elements inessential for us by the star. Here we
introduce the following way of ordering of the basis vectors: the
first $N$ vectors generate the kernel and the others $N$ vectors
generate its supplement.

Let  the following equation be valid under some values of the
parameters of ${\cal L}(v_{*})$:
\begin{equation}
P_{1} = \phi_{1} {\cal L}(u v_{*} \lambda), \qquad P_{2} =
\phi_{2} {\cal L}(u v_{*} \lambda^{-1}). \label{eq:PIP}
\end{equation}
Then after multiplying $n$ copies of the operator ${\cal L}_{1
3}(u v_{*}) {\cal R}_{1 2}(u)$ in the spaces '2' and '3' (see
Fig.2) and taking the trace we obtain the equation
\begin{equation}
\tilde{{\cal Q}}(u v_{*}) T(u) = \phi_{1}^{n} \tilde{{\cal Q}}(u
v_{*} \lambda) + \phi_{2}^{n} \tilde{{\cal Q}}(u v_{*}
\lambda^{-1}), \label{eq:TQ-ur}
\end{equation}
where
$$
\tilde{{\cal Q}}(u v_{*}) = tr_{3} \; {\cal L}_{1 3}(u v_{*}){\cal
L}_{1' 3}(u v_{*})...{\cal L}_{1^{(n)} 3},
$$
and $T(u)$ is the usual transfer matrix of the ice model. Making
the substitution
$$
{\cal Q}(u) = \tilde{{\cal Q}}(u v_{*}),
$$
we obtain the $T Q$-equation.

\begin{figure}
\begin{center}
\begin{picture}(100,120)(0,0)

\put(0,40){\line(1,0){100}} \put(0,90){\line(1,0){100}}
\put(0,93){\line(1,0){100}} \put(20,20){\line(0,1){90}}
\put(40,20){\line(0,1){90}} \put(80,20){\line(0,1){90}}
\put(-8,37){$2$} \put(-8,89){$3$} \put(56,65){$...$}
\put(18,10){$1$} \put(38,10){$1'$} \put(78,10){$1^{(n)}$}

\put(20,110){\vector(0,-1){10}} \put(20,110){\vector(0,-1){45}}
\put(20,110){\vector(0,-1){83}} \put(40,110){\vector(0,-1){10}}
\put(40,110){\vector(0,-1){45}} \put(40,110){\vector(0,-1){83}}
\put(80,110){\vector(0,-1){10}} \put(80,110){\vector(0,-1){45}}
\put(80,110){\vector(0,-1){83}}

\put(0,40){\vector(1,0){10}} \put(0,40){\vector(1,0){30}}
\put(0,40){\vector(1,0){93}} \put(0,90){\vector(1,0){10}}
\put(0,90){\vector(1,0){30}} \put(0,90){\vector(1,0){93}}
\put(0,93){\vector(1,0){10}} \put(0,93){\vector(1,0){30}}
\put(0,93){\vector(1,0){93}}

\put(3,98){$u v_{*}$} \put(10,45){$u$} \put(30,45){$u$}
\put(70,45){$u$} \put(63,98){$u v_{*}$}

\put(20,40){\oval(8,8)[tl]} \put(40,40){\oval(8,8)[tl]}
\put(80,40){\oval(8,8)[tl]} \put(20,93){\oval(8,8)[tl]}
\put(40,93){\oval(8,8)[tl]} \put(80,93){\oval(8,8)[tl]}

\end{picture}
\end{center}
\caption{}
\end{figure}

It turns out that if we use the definition (\ref{eq:e2}) for the
operator ${\cal L}(v)$, to make the calculations to the end is
impossible. This is the case because there is not any set of the
parameters such as the conditions (\ref{eq:PIP}) satisfy.

However, one can redefine the operator ${\cal L}(v)$ in order the
condition (\ref{eq:PIP}) and, consequently, the equation
(\ref{eq:TQ-ur}), to be valid \cite{kn:gnus2}. For that $p_{1}$
has to depend on the spectral parameter $v$ and $p_{2}$ has not to
change: $$ p_{1}(v) = \{ a_{1} v^{-1}, b_{1}, c_{1}, d_{1} v \},
\qquad p_{2}(v) = \{ a_{2}, b_{2}, c_{2}, d_{2} \}. $$ It is
shown in Appendix \ref{tq} that ${\cal Q}(u) = {\cal Q}(u,
p_{1}(u), p_{2})$ actually satisfies the $TQ$-equation: $$ {\cal
Q}(u)T(u) = (u-u^{-1})^{n}{\cal Q}(u \omega)+(u \omega - u^{-1}
\omega^{-1})^{n} {\cal Q}(u \omega^{-1}). $$ Here $T(u)$ is the
usual transfer matrix of the ice model.

\section{The elementary isomorphisms in the special case.}

We consider two representations of the algebra of $L$-operators:
$L(u,p_{1},p_{2})$ and $L(u,p_{2},p_{1})$. Let the parameters
$p_{1}$ and $p_{2}$ be such that these representations
are equivalent. We introduce the operator $G$ satisfying the
equation
\begin{equation}
\begin{array}{l}
G(Z)L(u,p_{1},p_{2}) = L(u,p_{2},p_{1}) G(Z).
\end{array}
\label{eq:117a}
\end{equation}
Here $G(Z)$ acts in $N$-dimensional space.

We now consider two tensor products of a pair of cyclic
representations: $L_1(u,p_1,p_2) L_2(u,p_3,p_4)$ and
$L_1(u,p_1,p_3) L_2(u,p_2,p_4)$ (pay attention to the permutation
$p_2 \leftrightarrow p_3$). Let the parameters $p_1$, $p_2$,
$p_3$, $p_4$ be such that the two representations are equivalent.
We introduce the operator $F$ intertwining them:
\begin{equation}
\begin{array}{l}
F(X_{1}X_{2}^{-1})L_{1}(u,p_{1},p_{2}) L_{2}(u,p_{3},p_{4})=\\
\vspace{-2mm}\\
=L_{1}(u,p_{1},p_{3}) L_{2}(u,p_{2},p_{4}) F(X_{1}X_{2}^{-1}).
\end{array}
\label{eq:118a}
\end{equation}
Here $F(X_1 X_2^{-1})$ acts in the $N$-dimensional space too.

It turns out that the conditions (\ref{eq:ququ}) are certainly
sufficient $G$ and $F$ to exist. If they obey, the two operators
are given by the following formulae:
\begin{equation}
\frac{G(p_{1},p_{2}; \omega^{k})}{G(p_{1},p_{2}; 1)}=
\prod_{j=1}^{k} \frac{d_{1}b_{2} -
a_{1}c_{2}\omega^{j}}{b_{1}d_{2} - c_{1}a_{2}\omega^{j}},
\label{eq:G}
\end{equation}
\begin{equation}
\frac{F(p_{1},p_{2}; \omega^{k})}{F(p_{1},p_{2}; 1)} =
\prod_{j=1}^{k} \frac{\omega a_{1}d_{2} - d_{1} a_{2}
\omega^{j}}{c_{1}b_{2} - b_{1}c_{2}\omega^{j}}. \label{eq:F}
\end{equation}
Here we denote by $G(p_{1},p_{2}; \omega^{k})$ and $F(p_{1},p_{2};
\omega^{k})$ the diagonal matrix elements of the $N$-dimensional
matrices $G(p_{1},p_{2}; Z)$ è $F(p_{1},p_{2}; X_{1}X_{2}^{-1})$
in the eigenbasis corresponding to each of them ($G$ and $F$
cannot be bringen to the diagonal form at the same time).

In the following section we calculate the two operators in more
general case, and now we note that the formulae (\ref{eq:G}),
(\ref{eq:F}) coincide with the formulae (\ref{eq:116}).

The existence of elementary isomorphisms $G$ and $F$ explains the
factorisation of the $R$-matrix of the chiral Potts model.
Permuting the pairs, we obtain the chain:
$$
(p_1,p_2)(p_3,p_4) \stackrel{F_1}{\rightarrow} (p_1,p_3)(p_2,p_4)
\stackrel{G_1,G_2}{\longrightarrow} (p_3,p_1)(p_4,p_2)
\stackrel{F_2}{\rightarrow} (p_3,p_4)(p_1,p_2).
$$

From this the factorisation of the $R$-matrix becomes evident.

\section{The general case.}

\subsection{The $G$-operator.}

Let us consider two representations of the algebra of
$L$-operators: $L(u,p_{1},\bar{p}_{1})$ and
$L(u,p_{2},\bar{p}_{2})$. We want to clarify when they are
equivalent and find the corresponding intertwiner which is 
a generalization of the operator $G$ introduced in the
previous part. For simplicity we denote this generalized
intertwiner by the same symbol $G$.

The two representations are equivalent if the following equations
are valid (Appendix \ref{C}):
\begin{equation}
\left\{
\begin{array}{l}
a_{1}^{N} \bar{a}_{1}^{N} = a_{2}^{N}\bar{a}_{2}^{N}, \qquad
b_{1}^{N} \bar{b}_{1}^{N} = b_{2}^{N}\bar{b}_{2}^{N},\\
\vspace{-2mm}\\
\displaystyle \frac{\bar{c}_{1}^{N}\bar{d}_{1}^{N}}
{\bar{a}_{1}^{N} \bar{b}_{1}^{N}} =
\frac{c_{2}^{N}d_{2}^{N}}{a_{2}^{N}b_{2}^{N}}, \qquad
\frac{\bar{c}_{2}^{N} \bar{d}_{2}^{N}}{\bar{a}_{2}^{N}
\bar{b}_{2}^{N}} =
\frac{c_{1}^{N}d_{1}^{N}}{a_{1}^{N}b_{1}^{N}},\\
\vspace{-2mm}\\
\displaystyle \frac{d_{1}^{N} \bar{d}_{1}^{N}}{a_{1}^{N}
\bar{a}_{1}^{N}} =
\frac{d_{2}^{N} \bar{d}_{2}^{N}}{a_{2}^{N} \bar{a}_{2}^{N}},\\
\vspace{-2mm}\\
d_{1}^{N} \bar{b}_{1}^{N} - a_{1}^{N} \bar{c}_{1}^{N} = d_{2}^{N}
\bar{b}_{2}^{N} - a_{2}^{N} \bar{c}_{2}^{N}.
\end{array}
\right.
\label{eq:20}
\end{equation}

We consider the simplest case when we extract the N-th roots by
the simple striking out the letter $N$. As the result we obtain
the system:
\begin{equation}
\left\{
\begin{array}{l}
a_{1} \bar{a}_{1} = a_{2} \bar{a}_{2}, \qquad
b_{1} \bar{b}_{1} = b_{2} \bar{b}_{2},\\
\vspace{-2mm}\\
\displaystyle \frac{\bar{c}_{1} \bar{d}_{1}} {\bar{a}_{1}
\bar{b}_{1}} = \frac{c_{2} d_{2}}{a_{2} b_{2}}, \qquad
\frac{\bar{c}_{2} \bar{d}_{2}}{\bar{a}_{2} \bar{b}_{2}} =
\frac{c_{1} d_{1}}{a_{1} b_{1}},\\
\vspace{-2mm}\\
d_{1} \bar{d}_{1} = d_{2} \bar{d}_{2},\\
\vspace{-2mm}\\
d_{1}^{N} \bar{b}_{1}^{N} - a_{1}^{N} \bar{c}_{1}^{N} = d_{2}^{N}
\bar{b}_{2}^{N} - a_{2}^{N} \bar{c}_{2}^{N}.
\end{array}
\right.
\label{eq:21}
\end{equation}

We find the operator $G$ satisfying the equation $$ G
L(u,p_{1},\bar{p}_{1}) = L(u,p_{2},\bar{p}_{2}) G. $$ If the
conditions (\ref{eq:21}) obey then the operator $G$ exists. We
prove this now.

Let us make use of the following ansaz:
$$
G = G(Z).
$$
We obtain the system:
$$
\left\{
\begin{array}{l}
G(Z)A_{1} = A_{2}G(Z),\\
\vspace{-2mm}\\
G(Z)B_{1} = B_{2}G(Z),\\
\vspace{-2mm}\\
G(Z)C_{1} = C_{2}G(Z),\\
\vspace{-2mm}\\
G(Z)D_{1} = D_{2}G(Z).
\end{array}
\right.
$$

We choose a basis $|k \rangle$, $k=0,...,N-1$ (mod $N$): $$ Z|k
\rangle = \omega^{k}|k \rangle, \qquad X |k \rangle = |k + 1
\rangle. $$ It is clear that in this basis the matrix $G(Z)$ is
diagonal. Let us find its nonzero matrix elements.

The first equation is
$$
G(Z)\left[ c_{1}\bar{c}_{1}Z - b_{1}\bar{b}_{1}u \right] = \left[
c_{2}\bar{c}_{2}Z - b_{2}\bar{b}_{2}u \right] G(Z).
$$

We act on the vector $|k \rangle$ by the left-hand and the
right-hand sides and compare the coefficients at different powers
of $u$. As the result we obtain the following restrictions on the
parameters:
\begin{equation}
c_{1} \bar{c}_{1} = c_{2} \bar{c}_{2}, \qquad  b_{1} \bar{b}_{1} =
b_{2} \bar{b}_{2}.
\label{eq:22}
\end{equation}

In the similar way, from the fourth equation,
$$
G(Z)\left[ d_{1}\bar{d}_{1} - a_{1}\bar{a}_{1}\omega u Z \right] =
\left[ d_{2} \bar{d}_{2} - a_{2} \bar{a}_{2} \omega u Z \right],
$$
we obtain that
\begin{equation}
d_{1}\bar{d}_{1}=d_{2}\bar{d}_{2}, \qquad
a_{1}\bar{a}_{1}=a_{2}\bar{a}_{2}.
\label{eq:23}
\end{equation}

The second equation is
$$
G(Z)X^{-1}\left[ d_{1}\bar{b}_{1} - a_{1}\bar{c}_{1}Z \right] =
\left[ d_{2}\bar{b}_{2} - a_{2}\bar{c}_{2}Z \right] G(Z).
$$
From this it follows that
\begin{equation}
G(\omega^{k+1}) =
\frac{d_{1}\bar{b}_{1}-a_{1}\bar{c}_{1}\omega^{k+1}}
{d_{2}\bar{b}_{2} - a_{2}\bar{c}_{2}\omega^{k+1}} G(\omega^{k}),
\label{eq:24}
\end{equation}
where by $G(\omega^{k})$, $k=0,...,N-1$, are denoted the diagonal
matrix elements of the matrix $G(Z)$.

In the same way from the third equation,
$$
G(Z)\left[ b_{1}\bar{d}_{1} - c_{1}\bar{a}_{1} Z \right]X = \left[
b_{2}\bar{d}_{2} - c_{2}\bar{a}_{2}Z \right]X G(Z),
$$
one can easily derive that
\begin{equation}
G(\omega^{k+1}) =
\frac{b_{2}\bar{d}_{2}-c_{2}\bar{a}_{2}\omega^{k+1}}
{b_{1}\bar{d}_{1}-c_{1}\bar{a}_{1}\omega^{k+1}} G(\omega^{k}).
\label{eq:25}
\end{equation}

Since $G(\omega^{k})$ has a single meaning it must be
$$
(d_{1}\bar{b}_{1}-a_{1}\bar{c}_{1}\omega^{k+1})
(b_{1}\bar{d}_{1}-c_{1}\bar{a}_{1}\omega^{k+1})=
(d_{2}\bar{b}_{2}-a_{2}\bar{c}_{2}\omega^{k+1})
(b_{2}\bar{d}_{2}-c_{2}\bar{a}_{2}\omega^{k+1}).
$$

Comparing coefficients at different powers of $\omega$ and taking
into account (\ref{eq:22}), (\ref{eq:23}), we obtain an
additional condition:
\begin{equation}
\frac{\bar{c}_{1}\bar{d}_{1}}{\bar{a}_{1}\bar{b}_{1}} +
\frac{c_{1}d_{1}}{a_{1}b_{1}} =
\frac{\bar{c}_{2}\bar{d}_{2}}{\bar{a}_{2}\bar{b}_{2}} +
\frac{c_{2}d_{2}}{a_{2}b_{2}}.
\label{eq:26}
\end{equation}

Apart from, we obtain from the periodicity condition
$G(\omega^{N+1})=G(\omega)$:
\begin{equation}
d_{1}^{N} \bar{b}_{1}^{N} - a_{1}^{N} \bar{c}_{1}^{N} = d_{2}^{N}
\bar{b}_{2}^{N} - a_{2}^{N} \bar{c}_{2}^{N}. \label{eq:27}
\end{equation}

Using the gauge symmetries of the $L$-operators, one can set
\begin{equation}
a_{1} = \bar{a}_{2}, \qquad b_{1}=\bar{b}_{2}.
\label{eq:28}
\end{equation}
Then from (\ref{eq:21}) it follows that (\ref{eq:27}),
(\ref{eq:22}), (\ref{eq:23}), (\ref{eq:26}) are valid.

Thus $G(\omega^{k})$ exists and is given by the recurrence
relation (\ref{eq:24}).

One can rewrite $G(\omega^{k})$ in terms of $p_{1}$, $\bar{p}_{1}$
without difficult. We substitute $$ \bar{c}_{2} =
\frac{c_{1}\bar{c}_{1}}{c_{2}} $$ into (\ref{eq:27}) and express
$c_{2}$ in terms of $p_{1}$, $\bar{p}_{1}$: $$ c_{2} = \bar{c}_{1}
\sqrt[N]{\frac{b_{1}^{N}\bar{d}_{1}^{N}-c_{1}^{N}\bar{a}_{1}^{N}}
{d_{1}^{N}\bar{b}_{1}^{N}-a_{1}^{N}\bar{c}_{1}^{N}}} =
\Lambda(p_1,\bar{p}_{1}) \bar{c}_{1}, $$ where we introduce a new
function $$
\Lambda(p_1,p_2)=\sqrt[N]{\frac{b_{1}^{N}\bar{d}_{1}^{N}-c_{1}^{N}\bar{a}_{1}^{N}}
{d_{1}^{N}\bar{b}_{1}^{N}-a_{1}^{N}\bar{c}_{1}^{N}}}. $$

From this one can obtain $$
\frac{G(p_{1},\bar{p}_{1};\omega^{k})}{G(p_{1},\bar{p}_{1};1)}
=\Lambda(p_1,\bar{p}_{1})^{k} \prod_{j=1}^{k}
\frac{d_{1}\bar{b}_{1} - a_{1} \bar{c}_{1} \omega^{j}}
{\bar{d}_{1} b_{1} - \bar{a}_{1} c_{1} \omega^{j}}. $$

We emphasize that $G$ depends only on $p_{1}$, $\bar{p}_{1}$.

So, the found operator $G$ generates an isomorphism of the two
representations of the algebra of monodromy matrices,
$L(u,p_{1},\bar{p}_{1})$, $L(u,p_{2},\bar{p}_{2})$, with the
parameters $p_{2}$, $\bar{p}_{2}$ expressing in terms of $p_{1}$,
$\bar{p}_{1}$ in the following way:
\begin{equation}
\begin{array}{l}
a_{2}=\bar{a}_{1}, \qquad \bar{a}_{2}=a_{1},\\
\vspace{-2mm}\\
b_{2}=\bar{b}_{1}, \qquad \bar{b}_{2}=b_{1},\\
\vspace{-2mm}\\
\displaystyle c_{2} = \Lambda(p_1,\bar{p}_{1}) \bar{c}_{1}, \qquad
\bar{c}_{2} = \Lambda(p_1,\bar{p}_{1})^{-1} c_{1},\\
\vspace{-2mm}\\
\displaystyle d_{2} = \Lambda(p_1,\bar{p}_{1})^{-1} \bar{d}_{1},
\qquad \bar{d}_{2} = \Lambda(p_1,\bar{p}_{1}) d_{1}.
\end{array}
\label{eq:29}
\end{equation}

The operator $G$ found here is a generalization of the operator
$G$ in (\ref{eq:116}). In order to come to such special case we
must set
$$
p_{2} = \bar{p}_{1}, \qquad \bar{p}_{2} = p_{1}.
$$
At that there is an additional constraint on the parameters
$p_{1}$, $\bar{p}_{1}$:
$$
\Lambda(p_1,\bar{p}_{1}) = 1.
$$

{\bf Note.} We extract the N-th roots by simple striking out the
letter $N$. But in all probability the general case comes to this
one. A complete investigation can be transacted by the following
way. In order two representations to be equivalent it is
necessary their centres coincide. However, when we derive our
conditions, we does not compare all central elements. We must add
the condition of equality of the corresponding quantum
determinants to our system of equations. We do not study
completely the question about what comes the additional condition
to. It seems that it can be used for investigating how one must
extract the $N$-th root, and it strongly restricts a number of
variants.

\subsection{The $F$-operator.}

Let us consider two representations of the algebra of
$L$-operators:
$L_{1}(u,p_{1},\bar{p}_{1})L_{2}(u,p_{2},\bar{p}_{2})$ and\\
$L_{1}(u,p_{3},\bar{p}_{3})L_{2}(u,p_{4},\bar{p}_{4})$. We want to
find the conditions under which these two representations are
equivalent and calculate the corresponding intertwiner. The
latter is a generalization of the operator $F$ introduced in the
previous section.

The matrix of the central elements:
\begin{equation}
\langle L(u,p,\bar{p}) \rangle = \left[
\begin{array}{ll}
c^{N} \bar{c}^{N} - b^{N} \bar{b}^{N}u & -u(b^{N} \bar{d}^{N} - c^{N} \bar{a}^{N})\\
\vspace{-2mm}\\
d^{N} \bar{b}^{N} - a^{N} \bar{c}^{N} & d^{N} \bar{d}^{N} - a^{N}
\bar{a}^{N} u
\end{array}
\right]. \label{eq:1}
\end{equation}

The necessary condition of the equivalence of the two
representations is the coincidence of their centres. Therefore $$
\langle L_{1}(u,p_{1},\bar{p}_{1}) \rangle \langle
L_{2}(u,p_{2},\bar{p}_{2}) \rangle = \langle
L_{1}(u,p_{3},\bar{p}_{3}) \rangle \langle
L_{2}(u,p_{4},\bar{p}_{4}) \rangle. $$ From this it follows that
\begin{equation}
\begin{array}{l}
\det \langle L_{1}(u,p_{1},\bar{p}_{1}) \rangle \cdot \det
\langle L_{2}(u,p_{2},\bar{p}_{2}) \rangle = \\
\vspace{-2mm}\\
\qquad \qquad \qquad = \det \langle L_{1}(u,p_{3},\bar{p}_{3})
\rangle \cdot \det \langle L_{2}(u,p_{4},\bar{p}_{4}) \rangle.
\end{array}
\label{eq:4}
\end{equation}

In Appendix \ref{C} it is shown that from these conditions one
can derive the following relations between the parameters (we do
not consider the trivial case $p_{3} = p_{1}$,
$\bar{p}_{3}=\bar{p}_{1}$, $p_{4}=p_{2}$,
$\bar{p}_{4}=\bar{p}_{2}$):
\begin{equation}
\left\{
\begin{array}{l}
b_{1}^{N} \bar{b}_{1}^{N} = b_{3}^{N} \bar{b}_{3}^{N}, \qquad
b_{2}^{N}
\bar{b}_{2}^{N} = b_{4}^{N} \bar{b}_{4}^{N},\\
\vspace{-2mm}\\
a_{1}^{N} \bar{a}_{1}^{N} = a_{3}^{N} \bar{a}_{3}^{N}, \qquad
a_{2}^{N}
\bar{a}_{2}^{N} = a_{4}^{N} \bar{a}_{4}^{N},\\
\vspace{-2mm}\\
d_{1}^{N} \bar{b}_{1}^{N} = d_{3}^{N} \bar{b}_{3}^{N}, \qquad
a_{2}^{N}
\bar{c}_{2}^{N} = a_{4}^{N} \bar{c}_{4}^{N},\\
\vspace{-2mm}\\
\displaystyle \frac{c_{2}^{N} d_{2}^{N}}{a_{2}^{N} b_{2}^{N}} =
\frac{\bar{c}_{3}^{N} \bar{d}_{3}^{N}}{\bar{a}_{3}^{N}
\bar{b}_{3}^{N}}, \qquad \frac{c_{1}^{N}d_{1}^{N}}{a_{1}^{N}
b_{1}^{N}} = \frac{c_{3}^{N}
d_{3}^{N}}{a_{3}^{N} b_{3}^{N}},\\
\vspace{-2mm}\\
\displaystyle \frac{\bar{c}_{2}^{N}
\bar{d}_{2}^{N}}{\bar{a}_{2}^{N} \bar{b}_{2}^{N}} =
\frac{\bar{c}_{4}^{N} \bar{d}_{4}^{N}}{\bar{a}_{4}^{N}
\bar{b}_{4}^{N}}, \qquad \frac{c_{4}^{N} d_{4}^{N}}{a_{4}^{N}
b_{4}^{N}} = \frac{\bar{c}_{1}^{N}\bar{d}_{1}^{N}}{\bar{a}_{1}^{N}
\bar{b}_{1}^{N}},\\
\vspace{-2mm}\\
\displaystyle b_{2}^{N} \bar{c}_{3}^{N} = c_{2}^{N}
\bar{b}_{1}^{N} \frac{b_{2}^{N}\bar{c}_{1}^{N} -
\bar{a}_{1}^{N}d_{2}^{N}}
{c_{2}^{N}\bar{b}_{1}^{N}-a_{2}^{N}\bar{d}_{1}^{N}},\\
\vspace{-2mm}\\
\displaystyle \bar{a}_{1}^{N} d_{4}^{N} = \bar{d}_{1}^{N}
a_{2}^{N} \frac{b_{2}^{N}\bar{c}_{1}^{N} -
\bar{a}_{1}^{N}d_{2}^{N}}
{c_{2}^{N}\bar{b}_{1}^{N}-a_{2}^{N}\bar{d}_{1}^{N}}.
\end{array}
\right.
\label{eq:6}
\end{equation}

We strike out the letter $N$ again and obtain:
\begin{equation}
\left\{
\begin{array}{l}
b_{1} \bar{b}_{1} = b_{3} \bar{b}_{3}, \qquad b_{2} \bar{b}_{2} =
b_{4} \bar{b}_{4},\\
\vspace{-2mm}\\
a_{1} \bar{a}_{1} = a_{3} \bar{a}_{3}, \qquad a_{2} \bar{a}_{2} =
a_{4} \bar{a}_{4},\\
\vspace{-2mm}\\
d_{1} \bar{b}_{1} = d_{3} \bar{b}_{3}, \qquad a_{2} \bar{c}_{2} =
\
a_{4} \bar{c}_{4},\\
\vspace{-2mm}\\
\displaystyle \frac{c_{2} d_{2}}{a_{2} b_{2}} =
\frac{\bar{c}_{3}\bar{d}_{3}} {\bar{a}_{3}\bar{b}_{3}}, \qquad
\frac{c_{1}d_{1}}{a_{1} b_{1}} =
\frac{c_{3}d_{3}}{a_{3} b_{3}},\\
\vspace{-2mm}\\
\displaystyle \frac{\bar{c}_{2} \bar{d}_{2}}{\bar{a}_{2}
\bar{b}_{2}} = \frac{\bar{c}_{4}\bar{d}_{4}}
{\bar{a}_{4}\bar{b}_{4}}, \qquad \frac{c_{4} d_{4}}{a_{4} b_{4}}
= \frac{\bar{c}_{1}\bar{d}_{1}}{\bar{a}_{1} \bar{b}_{1}}.
\end{array}
\right. \label{eq:7}
\end{equation}

Let us recall about the gauge symmetries of the $L$-operator. The
substitution
$$
a \rightarrow \lambda a, \qquad \bar{a} \rightarrow \lambda^{-1}
\bar{a},
$$
$$
c \rightarrow \lambda c, \qquad \bar{c} \rightarrow \lambda^{-1}
\bar{c},
$$
$$
b \rightarrow b, \qquad \bar{b} \rightarrow \bar{b}, \qquad d
\rightarrow d, \qquad \bar{d} \rightarrow \bar{d}
$$
does not change the operator $L(u,p,\bar{p})$. The same is valid
for the substitution
$$
b \rightarrow \lambda b, \qquad \bar{b} \rightarrow \lambda^{-1}
\bar{b},
$$
$$
d \rightarrow \lambda d, \qquad \bar{d} \rightarrow \lambda^{-1}
\bar{d},
$$
$$
a \rightarrow a, \qquad \bar{a} \rightarrow \bar{a}, \qquad c
\rightarrow c, \qquad \bar{c} \rightarrow \bar{c}.
$$

Using the gauge degrees of freedom, one can set
\begin{equation}
b_{1}=b_{3}, \qquad b_{2}=b_{4}, \qquad a_{1}=a_{3}, \qquad
a_{2}=a_{4}. \label{eq:8}
\end{equation}

From this and from (\ref{eq:7}) it follows that
\begin{equation}
\begin{array}{l}
\bar{b}_{1} = \bar{b}_{3}, \qquad \bar{b}_{2}=\bar{b}_{4}, \qquad
\bar{a}_{1}=\bar{a}_{3},
\qquad \bar{a}_{2}=\bar{a}_{4},\\
\vspace{-2mm}\\
d_{1}=d_{3}, \qquad \bar{d}_{2}=\bar{d}_{4}, \qquad c_{1}=c_{3},
\qquad \bar{c}_{2}=\bar{c}_{4}.
\end{array}
\label{eq:9}
\end{equation}

Apart from the equations satisfy:
\begin{equation}
\bar{c}_{3}\bar{d}_{3}=c_{2}d_{2}\frac{\bar{a}_{3}\bar{b}_{3}}{a_{2}b_{2}},
\qquad
c_{4}d_{4}=\bar{c}_{1}\bar{d}_{1}\frac{a_{2}b_{2}}{\bar{a}_{1}\bar{b}_{1}}.
\label{eq:10}
\end{equation}

Recalling the formulae (\ref{eq:6}), we obtain:
\begin{equation}
\bar{c}_{3}^{N} = \frac{c_{2}^{N}\bar{b}_{1}^{N}}{b_{2}^{N}}
\frac{b_{2}^{N}\bar{c}_{1}^{N} - \bar{a}_{1}^{N}d_{2}^{N}}
{c_{2}^{N}\bar{b}_{1}^{N}-a_{2}^{N}\bar{d}_{1}^{N}}, \label{eq:11}
\end{equation}
\begin{equation}
d_{4}^{N} = \frac{\bar{d}_{1}^{N}a_{2}^{N}}{\bar{a}_{1}^{N}}
\frac{b_{2}^{N}\bar{c}_{1}^{N} - \bar{a}_{1}^{N}d_{2}^{N}}
{c_{2}^{N}\bar{b}_{1}^{N}-a_{2}^{N}\bar{d}_{1}^{N}}. \label{eq:12}
\end{equation}

Now we want to find the matrix $F$ intertwining the two
representations in question. In particular, we want to prove that
the conditions (\ref{eq:8}) $-$ (\ref{eq:12}) are not only
necessary but sufficient for the existence of $F$.

So, we write
\begin{equation}
F L_{1}(p_{1},\bar{p}_{1}) L_{2}(p_{2},\bar{p}_{2}) =
L_{1}(p_{3},\bar{p}_{3}) L_{2}(p_{4},\bar{p}_{4}) F. \label{eq:13}
\end{equation}

We find the operator $F$ in the form $F(X_{1} X_{2}^{-1})$, where
$X_{1}$, $X_{2}$ are the matrix of shift acting in the first and
in the second $N$-dimensional spaces correspondingly.

As shown in Appendix \ref{D}, $F$ actually exists and is
expressed in terms of $p_{1}$, $\bar{p}_{1}$, $p_{2}$,
$\bar{p}_{2}$ by the formula:
$$
\frac{F(\bar{p}_{1},p_{2};\omega^{k})}{F(\bar{p}_{1},p_{2};1)}=
\Omega(\bar{p}_1,p_2)^{-k} \prod_{j=1}^{k} \frac{\bar{c}_{1}b_{2}
-\bar{a}_{1}d_{2}\omega^{j}}{\bar{b}_{1}c_{2}-
\bar{d}_{1}a_{2}\omega^{j}},
$$
where for the further convenience we define a new function
$$
\Omega(p_1,p_2) = \sqrt[N]{\frac{b_{2}^{N} \bar{c}_{1}^{N} -
d_{2}^{N} \bar{a}_{1}^{N} }{c_{2}^{N} \bar{b}_{1}^{N} -
a_{2}^{N}\bar{d}_{1}^{N}}}
$$

We emphasize that $F$ depends only on the parameters $\bar{p}_{1}$
and $p_{2}$.

At that the action of $F$ is given by the formulae
\begin{equation}
\begin{array}{l}
a_{3}=a_{1}, \qquad \bar{a}_{3}=\bar{a}_{1}, \qquad a_{4}=a_{2},
\qquad
\bar{a}_{4}=\bar{a}_{2},\\
\vspace{-2mm}\\
b_{3}=b_{1}, \qquad \bar{b}_{3}=\bar{b}_{1}, \qquad b_{4}=b_{2},
\qquad
\bar{b}_{4}=\bar{b}_{2},\\
\vspace{-2mm}\\
c_{3}=c_{1}, \qquad d_{3}=d_{1}, \qquad \bar{c}_{4}=\bar{c}_{2},
\qquad
\bar{d}_{4}=\bar{d}_{2},\\
\vspace{-2mm}\\
\displaystyle \bar{c}_{3} = \Omega(\bar{p}_1,p_2) \frac{c_{2}
\bar{b}_{1}}{b_{2}}, \qquad c_{4} = \Omega(\bar{p}_1,p_2)^{-1}
\frac{\bar{c}_{1} b_{2}}{\bar{b}_{1}}\\
\vspace{-2mm}\\
\displaystyle \bar{d}_{3} = \Omega(\bar{p}_1,p_2)^{-1} \frac{d_{2}
\bar{a}_{1}}{a_{2}}, \qquad d_{4} = \Omega(\bar{p}_1,p_2)
\frac{\bar{d}_{1}a_{2}}{\bar{a}_{1}},
\end{array}
\label{eq:15.5}
\end{equation}

It is clear that the obtained expression for $F$ is gauge
invariant.

Recall that the tensor product of two $L$-operators possesses another 
symmetry. Namely, one can insert the unity between the two factors:
$$
L_{1}(p_{3},\bar{p}_{3}) L_{2}(p_{4},\bar{p}_{4}) =
L_{1}(p_{3},\bar{p}_{3}) M^{-1} M L_{2}(p_{4},\bar{p}_{4}),
$$
where an arbitrary matrix $2 \times 2$ is denoted by $M$.

We set
$$
M = \left[
\begin{array}{ll}
\bar{b}_{1}/b_{2} & 0\\
\vspace{-2mm}\\
0 & \bar{a}_{1}/a_{2}
\end{array}
\right].
$$

We apply the additional symmetry to our $L$-operators. As the result,
we obtain the product of new $L$-operators, whose parameters are expressed 
in terms of $p_{1}$, $\bar{p}_{1}$, $p_{2}$, $\bar{p}_{2}$ in 
the following way:
\begin{equation}
\begin{array}{l}
a_{3}=a_{1}, \qquad \bar{a}_{3}=a_{2}, \qquad a_{4}=\bar{a}_{1},
\qquad
\bar{a}_{4}=\bar{a}_{2},\\
\vspace{-2mm}\\
b_{3}=b_{1}, \qquad \bar{b}_{3}=b_{2}, \qquad b_{4}=\bar{b}_{1},
\qquad
\bar{b}_{4}=\bar{b}_{2},\\
\vspace{-2mm}\\
c_{3}=c_{1}, \qquad d_{3}=d_{1}, \qquad \bar{c}_{4}=\bar{c}_{2},
\qquad
\bar{d}_{4}=\bar{d}_{2},\\
\vspace{-2mm}\\
\displaystyle \bar{c}_{3} =  \Omega(\bar{p}_1,p_2) c_{2},
\qquad c_{4} = \Omega(\bar{p}_1,p_2)^{-1} \bar{c}_{1},\\
\vspace{-2mm}\\
\displaystyle \bar{d}_{3}= \Omega(\bar{p}_1,p_2)^{-1} d_{2},
\qquad d_{4} = \Omega(\bar{p}_1,p_2) \bar{d}_{1}.
\end{array}
\label{eq:15.6}
\end{equation}

At that the expression for $F(\omega^{k+1})$, of course, is
invariable.

The case considered in the paper \cite{kn:gnus3}, is the special
case of the approach in question and can be obtained if we set in all
formulae
$$
p_{3} = p_{1}, \qquad \bar{p}_{3}=p_{2}, \qquad p_{4}=\bar{p}_{1},
\qquad \bar{p}_{4}=\bar{p}_{2}.
$$
In particular, it is not difficult to derive the $F$-operator in
(\ref{eq:116}). At that there is an additional constraint on the
parameters $\bar{p}_{1}$, $p_{2}$:
$$
\Omega(\bar{p}_1,p_2) = 1.
$$

The found operators $G$ and $F$ satisfy the relation that 
generalizes the star-triangle relation of the chiral Potts 
model \cite{kn:gnus1}. Namely, the equation is valid:
\begin{equation}
\begin{array}{l}
G(p,q; Z_{1}) F(\tilde{p},r;
X_{1}X_{2}^{-1}) G(\tilde{q},\tilde{r}; Z_{1}) =\\
\vspace{-2mm}\\
\mu F(q,r; X_{1}X_{2}^{-1}) G(p,r'; Z_{1}) F(p',q';
X_{1}X_{2}^{-1}),
\end{array}
\label{eq:a}
\end{equation}
where $\mu$ is a constant, and the intertwiners depend on the 
parameters that are expressed in terms of $p$, $q$, $r$ by 
the following formulae:
$$
\left\{
\begin{array}{l}
a'_r = a_{r},\\
\vspace{-2mm}\\
b'_r = b_{r},\\
\vspace{-2mm}\\
c'_r = \Omega(q,r) c_{r},\\
\vspace{-2mm}\\
d'_r = \Omega(q,r)^{-1} d_{r},
\end{array}
\right. \qquad \left\{
\begin{array}{l}
a'_{q} = a_{q},\\
\vspace{-2mm}\\
b'_{q} = b_{q},\\
\vspace{-2mm}\\
c'_{q} = \Omega(q,r)^{-1} c_{q},\\
\vspace{-2mm}\\
d'_{q} = \Omega(q,r) d_{q},
\end{array}
\right.
$$
$$
\left\{
\begin{array}{l}
a'_p = a_{p},\\
\vspace{-2mm}\\
b'_p = b_{p},\\
\vspace{-2mm}\\
c'_p = \Lambda(p,r')^{-1} c_{p},\\
\vspace{-2mm}\\
d'_p = \Lambda(p,r') d_{p},
\end{array}
\right. \qquad \left\{
\begin{array}{l}
\tilde{a}_{q} = a_{q},\\
\vspace{-2mm}\\
\tilde{b}_{q} = b_{q},\\
\vspace{-2mm}\\
\tilde{c}_{q} = \Lambda(p,q) c_{q},\\
\vspace{-2mm}\\
\tilde{d}_{q} = \Lambda(p,q)^{-1} d_{q},
\end{array}
\right.
$$
$$
\left\{
\begin{array}{l}
\tilde{a}_p = a_{p},\\
\vspace{-2mm}\\
\tilde{b}_p = b_{p},\\
\vspace{-2mm}\\
\tilde{c}_p = \Lambda(p,q)^{-1} c_{p},\\
\vspace{-2mm}\\
\tilde{d}_p = \Lambda(p,q) d_{p},
\end{array}
\right. \qquad \left\{
\begin{array}{l}
\tilde{a}_{r} = a_{r},\\
\vspace{-2mm}\\
\tilde{b}_{r} = b_{r},\\
\vspace{-2mm}\\
\tilde{c}_{r} = \Omega(\tilde{p},r) c_{r},\\
\vspace{-2mm}\\
\tilde{d}_{r} = \Omega(\tilde{p},r)^{-1} d_{r}.
\end{array}
\right.
$$

The proof of these statement one can find in Appendix \ref{A}.

{\bf Note.} The stated gives a foundation to say about the
existence of a new algebraic structure related to the cyclic
representations of the monodromy matrices algebra.

Let us consider a Hopf algebra with generators $L_i^j(p_1,p_2)$,
$i,j=0,1$, $p_1,p_2 \in {\bf C}^4$.

The coproduct is given by
$$
\Delta \left(L_{i}^{j}\right) = \left(L_{i}^{k}\right)_{1}
\left(L_{k}^{j}\right)_{2}.
$$

The relations in the Hopf algebra:
$$
\begin{array}{l}
G(p_{1},p_{2}) L_{i}^{j}(p_{1},p_{2}) =
L_{i}^{j}(\tilde{p}_{1},\tilde{p}_{2})
G(p_{1},p_{2}),\\
\vspace{-2mm}\\
F(\bar{p}_{1},p_{2}) \left(L_{i}^{k}(p_{1},\bar{p}_{1})
\right)_{1} \left(L_{k}^{j}(p_{2},\bar{p}_{2}) \right)_{2} =
\left(L_{i}^{k}(p_{1},\tilde{p}_{1}) \right)_{1}
\left(L_{k}^{j}(\tilde{p}_{2},\bar{p}_{2}) \right)_{2}
F(\bar{p}_{1},p_{2}),\\
\vspace{-2mm}\\
\left(L_{i}^{k}(p_{1},\bar{p}_{1}) \right)_{1}
\left(L_{k}^{j}(p_{2},\bar{p}_{2}) \right)_{2} =
\left(L_{i}^{k}(p_{1},\bar{p}_{1}) \right)_{1} M_{k}^{l}
(M^{-1})_{l}^{m} \left(L_{m}^{j}(p_{2},\bar{p}_{2}) \right)_{2},
\end{array}
$$
where an arbitrary two-dimensional diagonal matrix is denoted by $M$, 
and the parameters in the right-hand sides of the relations are
expressed in terms of the parameters in the left-hand sides by
the formulae (\ref{eq:29}), (\ref{eq:15.6}). We mean the sum over
repeating indices.

\section{The algebra of the $Q$-operators.}

Besides the operator ${\cal Q}(u)$ introduced in the Section 3
and related to the cyclic representations of the algebra of
monodromy matrices in the form by Bazhanov--Stroganov, one can
consider the operator $Q(u)$,
$$
Q(u) = {\rm tr}_{0} L_{1 0}(u)L_{2 0}(u)...L_{k 0}(u),
$$
where the trace is calculated in the $N$-dimensional space and 
the cyclic representations of the monodromy matrices algebra in 
the form by Tarasov are denoted by $L_{i 0}(u)$.

The operators $Q(u)$ generate the algebra with the relations 
following from the properties of the operators $L(u)$:
\begin{equation}
Q(\lambda p_{1},\bar{p}_{1}) = \lambda^{k} Q(p_{1},\bar{p}_{1}),
\label{eq:Sootn0a}
\end{equation}
\begin{equation}
Q(p_{1},\mu \bar{p}_{1}) = \mu^{k} Q(p_{1},\bar{p}_{1}),
\label{eq:Sootn0b}
\end{equation}
\begin{equation}
Q(a_{1},b_{1},c_{1},d_{1},\bar{a}_{1},
\bar{b}_{1},\bar{c}_{1},\bar{d}_{1}) = Q(\lambda
a_{1},b_{1},\lambda c_{1},d_{1},\lambda^{-1}\bar{a}_{1},
\bar{b}_{1},\lambda^{-1}\bar{c}_{1},\bar{d}_{1}),
\label{eq:Sootn3a}
\end{equation}
\begin{equation}
Q(a_{1},b_{1},c_{1},d_{1},\bar{a}_{1},
\bar{b}_{1},\bar{c}_{1},\bar{d}_{1}) = Q(a_{1},\mu b_{1},c_{1},\mu
d_{1},\bar{a}_{1},\mu^{-1} \bar{b}_{1},\bar{c}_{1},\mu^{-1}
\bar{d}_{1}),
\label{eq:Sootn3b}
\end{equation}
\begin{equation}
\begin{array}{l}
Q(a_{1},b_{1},c_{1},d_{1},\bar{a}_{1},
\bar{b}_{1},\bar{c}_{1},\bar{d}_{1}) = \qquad \qquad \\
\vspace{-2mm}\\
\qquad \qquad = Q(\bar{a}_{1},\bar{b}_{1},\Lambda \bar{c}_{1},
\Lambda^{-1} \bar{d}_{1}, a_{1},b_{1}, \Lambda^{-1} c_{1},\Lambda
d_{1}),
\end{array}
\label{eq:Sootn1}
\end{equation}
\begin{equation}
\begin{array}{l}
Q(a_{1},b_{1},c_{1},d_{1},\bar{a}_{1},
\bar{b}_{1},\bar{c}_{1},\bar{d}_{1})
Q(a_{2},b_{2},c_{2},d_{2},\bar{a}_{2},
\bar{b}_{2},\bar{c}_{2},\bar{d}_{2}) = \\
\vspace{-2mm}\\
\qquad \qquad \qquad =Q(a_{1},b_{1},c_{1},d_{1},\beta \bar{a}_{1},
\alpha \bar{b}_{1},\alpha \bar{c}_{1},\beta \bar{d}_{1})\\
\vspace{-2mm}\\
\qquad \qquad \qquad \qquad \times
Q(\beta^{-1}a_{2},\alpha^{-1}b_{2},\alpha^{-1}c_{2},\beta^{-1}
d_{2},\bar{a}_{2},\bar{b}_{2},\bar{c}_{2},\bar{d}_{2}),
\end{array}
\label{eq:Sootn2}
\end{equation}
\begin{equation}
\begin{array}{l}
Q(a_{1},b_{1},c_{1},d_{1},\bar{a}_{1},
\bar{b}_{1},\bar{c}_{1},\bar{d}_{1})
Q(a_{2},b_{2},c_{2},d_{2},\bar{a}_{2},
\bar{b}_{2},\bar{c}_{2},\bar{d}_{2}) =\\
\vspace{-2mm}\\
\qquad \qquad Q(a_{1},b_{1},c_{1},d_{1},a_{2},b_{2},\Omega c_{2},
\Omega^{-1} d_{2})\\
\vspace{-2mm}\\ \qquad \qquad \qquad \qquad \times
Q(\bar{a}_{1},\bar{b}_{1},
\Omega^{-1}\bar{c}_{1},\Omega\bar{d}_{1},
\bar{a}_{2},\bar{b}_{2},\bar{c}_{2},\bar{d}_{2}),
\end{array}
\label{eq:Sootn4}
\end{equation}
where $\alpha$, $\beta$, $\lambda$, $\mu$ are arbitrary numbers,
$$
\Lambda = \Lambda(p_{1},\bar{p}_{1}),  \qquad \Omega =
\Omega(\bar{p}_1,p_2).
$$

One can find the proof of all these relations in Appendix
\ref{AlQ}.

\section{Discussion.}

We note some questions which can become the topics of the future
investigations.

The approach used in this paper can be applied in more general
cases, namely it would be interesting to study the elementary
isomorphisms intertwining the cyclic representations of the
monodromy matrices algebra related to the elliptical
$R$-matrix and also the $R$-matrix corresponding to the quantum
algebra $U_{q}(sl_{n})$.

The spectrum of the six-vertex model transfer matrix in
roots of unity is degenerate \cite{kn:gnus2}, \cite{kn:McCoy1}.
Some finite dimensional representations of the $Q$-operators algebra 
correspond to non-one-dimensional eigensubspaces of the transfer matrix.
The $Q$-operators act on these spaces in nontrivial way,
since they do not generally commute with each other.

Therefore an investigation of finite dimensional representations
of the $Q$-operators algebra given by (\ref{eq:Sootn0a}) $-$
(\ref{eq:Sootn4}) can shed light on the properties of the 
transfer matrix spectrum.

It is also interesting to clarify the relationship between the
algebra of $Q$-operators and the $U(A_{1}^{1})$ symmetry found in
the paper \cite{kn:McCoy1}.

\section*{Acknowledgements.}

We are grateful to B.Feigin and Yu.Stroganov for useful
discussions. One of us (A.O.) is grateful to V.Bazhanov for the
discussion of the factorization of the chiral Potts model $R$-matrix.

This work is supported by the following grants: RBRF-01-02-16686,
RBRF-99-01-01169, RBRF-00-15-96579, CRDF-RP1-2254, INTAS-00-00055.

\section*{Appendices.}

\appendix

\section{$TQ$-equation.}
\label{tq}

So, we want to find the kernel of the operator, ${\cal L}_{2 3}(v;
p_{1}(v), p_{2}) = {\tilde{\cal L}}_{2 3}(v)$, where
$$
p_{1}(v) = \{ a_{1}v^{-1}, b_{1}, c_{1}, d_{1}v \}, \qquad p_{2}
= \{ a_{2}, b_{2}, c_{2}, d_{2} \}.
$$
We have
$$
{\tilde{\cal L}}(v) = \left[
\begin{array}{ll}
v^{-1}c_{1}c_{2}Z - b_{1}b_{2} v Z^{-1} &
-v(b_{1}d_{2}Z^{-1} - c_{1}a_{2}Z)X \\
\vspace{-2mm}\\
\omega v^{-1} X^{-1} (d_{1}b_{2}v Z^{-1} - a_{1}c_{2}v^{-1}Z) &
d_{1}d_{2}Z^{-1} - a_{1}a_{2} \omega^{2} Z
\end{array}
\right].
$$

Hereinafter we use the following basis $|\alpha\rangle$, $\alpha
= 0,..., N-1 (mod \; N)$:
$$
Z |\alpha \rangle = \omega^{\alpha} |\alpha \rangle, \qquad X
|\alpha \rangle = | \alpha + 1 \rangle.
$$
Let us consider $2N$-dimensional vector $\Psi$:
$$
\Psi = \left(
\begin{array}{l}
\Phi_{1}\\
\vspace{-2mm}\\
\Phi_{2}
\end{array}
\right),
$$
where $\Phi_{1}$, $\Phi_{2}$ are $N$-dimensional vectors.

We act on $\Psi$ by the operator ${\tilde{\cal L}}_{2 3}(v)$ and
compare the result with zero. We obtain:
\begin{equation}
\left\{
\begin{array}{l}
(c_{1}c_{2}Z-b_{1}b_{2}v^{2}Z^{-1})\Phi_{1}-
v^{2}(b_{1}d_{2}Z^{-1}-c_{1}a_{2}Z)X \Phi_{2}=0,\\
\vspace{-2mm}\\
\omega X^{-1}v^{-1}(d_{1}b_{2}v Z^{-1}-a_{1}c_{2}v^{-1}Z)\Phi_{1}
+ (d_{1}d_{2}Z^{-1} - a_{1}a_{2}\omega^{2}Z)\Phi_{2}=0.
\end{array}
\right.
\label{eq:tq1}
\end{equation}
It is not difficult to see that this system has a solution if and
only if
$$
v^{2} = v_{*}^{2} = \frac{c_{2}d_{2}}{a_{2} b_{2}}.
$$
The vectors generating the kernel are
$$
\Psi_{\alpha}= \left(
\begin{array}{l}
-d_{2}|\alpha \rangle\\
\vspace{-2mm}\\
 b_{2}|\alpha - 1 \rangle
\end{array}
\right) = - d_{2}|0,\alpha \rangle + b_{2}|1,\alpha - 1 \rangle.
$$

Now we must act on the vectors $\Psi_{\alpha}$ by the operator
${\tilde{\cal L}}_{1 3}(u v_{*}){\cal R}_{1 2}(u)$ (it is
especially worth to note that here we can employ the usual matrix
multiplication in the two-dimensional space).

We write
$$
{\tilde{\cal L}}_{1 3}(u v_{*}) = \left[
\begin{array}{ll}
A(u v_{*}) & B(u v_{*})\\
\vspace{-2mm}\\
C(u v_{*}) & D(u v_{*})
\end{array}
\right], \qquad {\cal R}_{1 2}(u) = \left[
\begin{array}{ll}
a(u) & b(u)\\
\vspace{-2mm}\\
c(u) & d(u)
\end{array}
\right].
$$

Then
$$
{\tilde{\cal L}}_{1 3}(u v_{*}) {\cal R}_{1 2}(u)= \left[
\begin{array}{ll}
a(u)A(u v_{*})+c(u)B(u v_{*}) & b(u)A(u v_{*})+d(u)B(u v_{*})\\
\vspace{-2mm}\\
a(u)C(u v_{*})+c(u)D(u v_{*}) & b(u)C(u v_{*})+d(u)D(u v_{*})
\end{array}
\right].
$$

Acting on the vector $\Psi_{\alpha}$ by each of the four matrix elements, 
we obtain, for example,
$$
({\tilde{\cal L}} {\cal R})_{0}^{0} \Psi_{\alpha}= [a(u)A(u
v_{*})+c(u)B(u v_{*})][-d_{2}|0,\alpha \rangle+ b_{2}|1,\alpha - 1
\rangle]=
$$
$$
\qquad = -d_{2}(u \omega - u^{-1} \omega^{-1}) (v_{*}^{-1} u^{-1}
c_{1}c_{2} \omega^{\alpha} - b_{1}b_{2} u v_{*}
\omega^{-\alpha})|0,\alpha \rangle +
$$
$$
+ \omega b_{2} \left\{ (u-u^{-1})(v_{*}^{-1} u^{-1} c_{1}c_{2}
\omega^{\alpha-1} - b_{1}b_{2} u v_{*}
\omega^{1-\alpha})|1,\alpha - 1 \rangle + \right.
$$
$$
+ \left. u(\omega - \omega^{-1})(-u v_{*}) (b_{1}d_{2}
\omega^{-\alpha} - c_{1}a_{2} \omega^{\alpha})|0,\alpha \rangle
\right\}=
$$
$$
=(u - u^{-1})(v_{*}^{-1} u^{-1} c_{1}c_{2} \omega^{\alpha-1} -
b_{1} b_{2} v_{*} u \omega^{1-\alpha}) \Psi_{\alpha}.
$$

Similarly,
$$
({\tilde{\cal L}} {\cal R})_{1}^{1} \Psi_{\alpha}=
(u-u^{-1})(d_{1}d_{2} \omega^{-\alpha} -
a_{1}a_{2}\omega^{2+\alpha})\Psi_{\alpha},
$$
$$
({\tilde{\cal L}} {\cal R})_{0}^{1} \Psi_{\alpha}=
(u-u^{-1})(-v_{*} u)(b_{1}d_{2} \omega^{-\alpha-1} -
c_{1}a_{2}\omega^{\alpha+1}) \Psi_{\alpha+1},
$$
$$
({\tilde{\cal L}} {\cal R})_{1}^{0} \Psi_{\alpha} = (u-u^{-1})
\omega v_{*}^{-1} u^{-1} (d_{1}b_{2}u v_{*} \omega^{1-\alpha} -
a_{1}c_{2}u^{-1}v_{*}^{-1} \omega^{\alpha-1}) \Psi_{\alpha-1}.
$$

It is clear that the kernel ${\tilde{\cal L}}_{2 3}(v_{*})$ is
actually an invariant subspace with regard to the operator
${\tilde{\cal L}}_{1 3}(u v_{*}) {\cal R}_{1 2}(u)$.

It is not difficult to see that the obtained $N$-dimensional
matrix is proportional to ${\tilde{\cal L}}(u v_{*} \omega)$:
$$
\left. {\tilde{\cal L}}_{1 3}(u v_{*}){\cal R}_{1 2}(u)
\right|_{Ker} = (u - u^{-1}) {\tilde{\cal L}}(u v_{*} \omega).
$$

Let us now consider the vectors that belong to the supplement of the
kernel. If we make a factorization by the vectors of the kernel
then this supplement must be an invariant space with regard to the
operator ${\tilde{\cal L}}_{1 3}(u v_{*}) {\cal R}_{1 2}(u)$. We
choose the basis
$$
\Psi'_{\alpha} = |0, \alpha \rangle.
$$
Making all computations we find ($mod \; \Psi_{\alpha}$):
$$
({\tilde{\cal L}} {\cal R})_{0}^{0} \Psi'_{\alpha}= (u \omega -
u^{-1} \omega^{-1}) (v_{*}^{-1} u^{-1} c_{1}c_{2} \omega^{\alpha}
- b_{1} b_{2} v_{*} u \omega^{-\alpha}) \Psi'_{\alpha}.
$$
$$
({\tilde{\cal L}} {\cal R})_{1}^{1} \Psi'_{\alpha}= (u \omega -
u^{-1} \omega^{-1}) (d_{1}d_{2} \omega^{1-\alpha} - a_{1}a_{2}
\omega^{1+\alpha})\Psi'_{\alpha},
$$
$$
({\tilde{\cal L}} {\cal R})_{0}^{1} \Psi'_{\alpha}= (u \omega -
u^{-1} \omega^{-1})(-v_{*} u)(b_{1}d_{2} \omega^{-\alpha} -
c_{1}a_{2}\omega^{\alpha}) \Psi'_{\alpha+1},
$$
$$
({\tilde{\cal L}} {\cal R})_{1}^{0} \Psi'_{\alpha} = (u \omega -
u^{-1} \omega^{-1}) \omega v_{*}^{-1} u^{-1} (d_{1}b_{2} u v_{*}
\omega^{-\alpha} - a_{1}c_{2} u^{-1}v_{*}^{-1} \omega^{\alpha})
\Psi'_{\alpha-1}.
$$

As in the previous case, it is not difficult to see that the found
$N$-dimensional matrix is proportional to ${\tilde{\cal L}}(u
v_{*} \omega^{-1})$:
$$
\left. {\tilde{\cal L}}_{1 3}(u v_{*}){\cal R}_{1 2}(u)
\right|_{Ker^{\perp}} = (u \omega - u^{-1} \omega^{-1})
{\tilde{\cal L}}(u v_{*} \omega^{-1}).
$$

Thus we obtain that the conditions (\ref{eq:PIP}) actually
satisfy. At that
$$
\lambda = \omega, \qquad \phi_{1} = u-u^{-1}, \qquad \phi_{2} = u
\omega - u^{-1} \omega^{-1}.
$$

As the result, the $TQ$-equation is valid:
$$
{\cal Q}(u)T(u) = (u-u^{-1})^{n}{\cal Q}(u \omega)+(u \omega -
u^{-1} \omega^{-1})^{n} {\cal Q}(u \omega^{-1}),
$$
where
$$
{\cal Q}(u) = tr_{3} \; {\tilde{\cal L}}_{1 3}(u) {\tilde{\cal
L}}_{1' 3}(u)... {\tilde{\cal L}}_{1^{(n)} 3}(u).
$$

\section{The conditions of the equivalence of representations.
The Ferma curve.}
\label{Ferma}

Let $N$ be odd.

We now prove that the representations
$L_{1}(u,p_{1},p_{2})L_{2}(u,p_{3},p_{4})$ and
$L_{1}(u,p_{3},p_{4})L_{2}(u,p_{1},p_{2})$ are equivalent in the
general case if and only if we can choose $p_{i}$, $i=1,2,3,4$,
satisfying
\begin{equation}
\frac{a_{i}^{N} \pm b_{i}^{N}}{c_{i}^{N} \pm d_{i}^{N}} =
\lambda_{\pm}. \label{eqz:2}
\end{equation}

For the convenience we make the following substitution
$$
a_{i}^{N} \rightarrow a_{i}, \qquad b_{i}^{N} \rightarrow b_{i},
\qquad c_{i}^{N} \rightarrow c_{i}, \qquad d_{i}^{N} \rightarrow
d_{i}.
$$

We have
\begin{equation}
\langle L(u,p_{1},p_{2}) \rangle = \left[
\begin{array}{ll}
c_{1} c_{2} - b_{1} b_{2} u & -u(b_{1}d_{2}-c_{1}a_{2})\\
\vspace{-2mm}\\
d_{1} b_{2} - a_{1} c_{2}   & d_{1}d_{2} - a_{1} a_{2}u
\end{array}
\right], \label{eqz:3}
\end{equation}

\begin{equation}
\langle L(u,p_{3},p_{4}) \rangle = \left[
\begin{array}{ll}
c_{3} c_{4} - b_{3} b_{4} u & -u(b_{3}d_{4}-c_{3}a_{4})\\
\vspace{-2mm}\\
d_{3} b_{4} - a_{3} c_{4}   & d_{3}d_{4} - a_{3} a_{4}u
\end{array}
\right]. \label{eqz:4}
\end{equation}

Two representations $L_{\pi}$, $L_{\pi'}$ are equivalent if there
exists an isomorphism $P$:
$$
L_{\pi'} = P L_{\pi} P^{-1},
$$
that is, elements of $L_{\pi}$ and $L_{\pi'}$ are in one-to-one
correspondence.

The central elements $\langle L_{\pi} \rangle$ and $\langle
L_{\pi'} \rangle$ must coincide \cite{kn:gnus3}.

If $\pi = \pi_{1} \times \pi_{2}$, $\pi' = \pi_{2} \times
\pi_{1}$, then
$$
\langle L_{\pi_{1}} \rangle \langle L_{\pi_{2}} \rangle = \langle
L_{\pi_{2}} \rangle \langle L_{\pi_{1}} \rangle.
$$

From this it follows that
$$
\langle L(u,p_{1},p_{2}) \rangle \langle L(u,p_{3},p_{4}) \rangle
= \langle L(u,p_{3},p_{4}) \rangle \langle L(u,p_{1},p_{2})
\rangle.
$$

Multiplying the matrices we obtain five equations. Only three of them 
are independent:
\begin{equation}
\frac{b_{1}d_{2}-c_{1}a_{2}}{d_{1} b_{2}-a_{1}c_{2}} =
\frac{b_{3}d_{4}-c_{3}a_{4}}{d_{3} b_{4}-a_{3}c_{4}} = s,
\label{eqz:5}
\end{equation}
\begin{equation}
\frac{a_{1}a_{2}-b_{1}b_{2}}{b_{1}d_{2}-c_{1}a_{2}} =
\frac{a_{3}a_{4}-b_{3}b_{4}}{b_{3}d_{4}-c_{3}a_{4}} = q,
\label{eqz:6}
\end{equation}
\begin{equation}
\frac{c_{1}c_{2}-d_{1}d_{2}}{b_{1}d_{2}-c_{1}a_{2}} =
\frac{c_{3}c_{4}-d_{3}d_{4}}{b_{3}d_{4}-c_{3}a_{4}} = r,
\label{eqz:7}
\end{equation}
where $s$, $q$, $r$ are arbitrary constants.

We want to find the constraints on $p_{i}$ under which 
this system has solutions. We have:
\begin{equation}
\left\{
\begin{array}{l}
b_{1}d_{2}-c_{1}a_{2}=s(d_{1}b_{2}-a_{1}c_{2}),\\
\vspace{-2mm}\\
a_{1}a_{2}-b_{1}b_{2}=q(b_{1}d_{2}-c_{1}a_{2}),\\
\vspace{-2mm}\\
c_{1}c_{2}-d_{1}d_{2}=r(b_{1}d_{2}-c_{1}a_{2}).
\end{array}
\right. \label{eqz:8}
\end{equation}

It turns out that (\ref{eqz:8}) can satisfy if $p_{1}$ and $p_{2}$
are points of the curve obtained by crossing of two planes (the
projective symmetry of the operator $L$):
$$
\left\{
\begin{array}{l}
\alpha_{1}a_{i} + \beta_{1}b_{i} + \gamma_{1}c_{i} +
\delta_{1}d_{i} = 0\\
\vspace{-2mm}\\
\alpha_{2}a_{i} + \beta_{2}b_{i} + \gamma_{2}c_{i} +
\delta_{2}d_{i} = 0
\end{array}
\right. ,\qquad i=1,2.
$$

Let us find these planes. We have from the last system:
\begin{equation}
\left\{
\begin{array}{l}
a_{i} = \lambda_{i}c_{i} + \mu_{i}d_{i},\\
\vspace{-2mm}\\
b_{i} = \nu_{i}c_{i} + \eta_{i}d_{i}
\end{array}
\right. , \qquad i=1,2. \label{eqz:9}
\end{equation}

We substitute (\ref{eqz:9}) into (\ref{eqz:8}) and obtain the
equations for coefficients:
$$
\left\{
\begin{array}{l}
\eta_{1} = s \eta_{2},\\
\vspace{-2mm}\\
\lambda_{2} = s \lambda_{1},\\
\vspace{-2mm}\\
\nu_{1}=\mu_{2},\\
\vspace{-2mm}\\
\nu_{2}=\mu_{1},\\
\vspace{-2mm}\\
1=-r \lambda_{2},\\
\vspace{-2mm}\\
-1 = r \eta_{1},\\
\vspace{-2mm}\\
\nu_{1}=\mu_{2},\\
\vspace{-2mm}\\
\lambda_{1} \lambda_{2} - \nu_{1} \nu_{2} = -q \lambda_{2},\\
\vspace{-2mm}\\
\mu_{1}\mu_{2}-\eta_{1}\eta_{2} = q \eta_{1},\\
\vspace{-2mm}\\
\lambda_{1} \mu_{2}-\nu_{1} \eta_{2} = q \nu_{1} - q \mu_{2},\\
\vspace{-2mm}\\
\mu_{1} \lambda_{2} - \nu_{2} \eta_{1} = 0.
\end{array}
\right.
$$

Solving the system, we find:
$$
\left\{
\begin{array}{l}
\nu_{1}=\mu_{2}=\mu,\\
\vspace{-2mm}\\
\mu_{1}=\nu_{2}=\nu,\\
\vspace{-2mm}\\
\eta_{1}=\lambda_{2}=\lambda,\\
\vspace{-2mm}\\
\lambda_{1}=\eta_{2}=\eta,\\
\vspace{-2mm}\\
\lambda=-1/r,\\
\vspace{-2mm}\\
\nu \mu = (q+\eta)\lambda,\\
\vspace{-2mm}\\
s=1.
\end{array}
\right.
$$

Apart from, since the points $p_{1}$ and $p_{2}$ lie on the same
curve, it must be
$$
\lambda_{1}=\lambda_{2}=\lambda, \qquad \mu_{1}=\mu_{2}=\mu,
\qquad \nu_{1}=\nu_{2}=\nu, \qquad \eta_{1}=\eta_{2}=\eta.
$$
Consequently,
$$
\lambda = \eta, \qquad \nu = \mu.
$$

As the result,
$$
\frac{a_{1} + b_{1}}{c_{1} + d_{1}} = \frac{a_{2} + b_{2}}{c_{2} +
d_{2}}, \qquad \frac{a_{1} - b_{1}}{c_{1} - d_{1}}=\frac{a_{2} -
b_{2}}{c_{2} - d_{2}}.
$$
Similarly, it must be
$$
\frac{a_{3} + b_{3}}{c_{3} + d_{3}} = \frac{a_{4} + b_{4}}{c_{4}
+ d_{4}}, \qquad \frac{a_{3} - b_{3}}{c_{3} - d_{3}}=\frac{a_{4}
- b_{4}}{c_{4} - d_{4}}.
$$

We return to the old notations:
$$
p_{i} \rightarrow p_{i}^{N}, \qquad i=1,2,3,4.
$$

In principle, the points $p_{1}$, $p_{2}$, $p_{3}$, $p_{4}$ can
be related in two different ways:
$$
1) \, \frac{a_{1}^{N} \pm b_{1}^{N}}{c_{1}^{N} \pm d_{1}^{N}} =
\frac{a_{2}^{N} \pm b_{2}^{N}}{c_{2}^{N} \pm d_{2}^{N}} =
\frac{a_{3}^{N} \pm b_{3}^{N}}{c_{3}^{N} \pm d_{3}^{N}} =
\frac{a_{4}^{N} \pm b_{4}^{N}}{c_{4}^{N} \pm d_{4}^{N}},
$$
$$
2) \, \frac{a_{1}^{N} \pm b_{1}^{N}}{c_{1}^{N} \pm d_{1}^{N}} =
\frac{a_{2}^{N} \pm b_{2}^{N}}{c_{2}^{N} \pm d_{2}^{N}} =
\frac{a_{3}^{N} \mp b_{3}^{N}}{c_{3}^{N} \mp d_{3}^{N}} =
\frac{a_{4}^{N} \mp b_{4}^{N}}{c_{4}^{N} \mp d_{4}^{N}}.
$$
This can be explained by the existence of two roots of the equation
$$
\nu^{2}=\mu^{2}=-q/r + 1/r^{2}.
$$

However, one should recall about symmetries of the operator
$L(u,p_{1},p_{2})$: one can substitute
$$
b_{1} \rightarrow \lambda b_{1}, \qquad b_{2} \rightarrow
\lambda^{-1} b_{2}, \qquad d_{1} \rightarrow \lambda d_{1},
\qquad d_{2} \rightarrow \lambda^{-1} d_{2},
$$
at that $L(u,p_{1},p_{2})$ does not change. With the aim of such
substitution, where $\lambda=-1$, the curve $2)$ reduces to the
curve $1)$ (recall that $N$ is odd).

\section{The conditions of equivalence of representations in the
general case.}
\label{C}

We introduce the notations:
\begin{equation}
\begin{array}{l}
\phi = d^{N} \bar{b}^{N}, \qquad \psi = a^{N} \bar{c}^{N},
\qquad \beta = b^{N} \bar{b}^{N},\\
\vspace{-2mm}\\
\displaystyle \delta = a^{N} \bar{a}^{N}, \qquad \mu = \frac{c^{N}
d^{N}}{a^{N} b^{N}}, \qquad \lambda = \frac{\bar{c}^{N}
\bar{d}^{N}}{\bar{a}^{N} \bar{b}^{N}}.
\end{array}
\label{eq:2}
\end{equation}

Then $\langle L(u,p,\bar{p}) \rangle$ can be written in the form
\begin{equation}
\langle L(u,p,\bar{p}) \rangle = \left[
\begin{array}{ll}
\displaystyle \frac{\mu \beta \psi}{\phi} - \beta u &
\displaystyle -u\left( \frac{\lambda
\beta \delta}{\psi} - \frac{\mu \beta \delta}{\phi} \right)\\
\vspace{-2mm}\\
\displaystyle \phi-\psi & \displaystyle \frac{\lambda \delta
\phi}{\psi} - \delta u
\end{array}
\right].
\label{eq:3}
\end{equation}

We consider two representations of the algebra of $L$-operators,
$L(u,p_{1},\bar{p}_{1})$ è $L(u,p_{2},\bar{p}_{2})$, and want to
find the conditions under which they are equivalent.

The necessary condition of the equivalence of the two
representations is a coincidence of centres of these
representations. We have:
\begin{equation}
\langle L(u,p_{1},\bar{p}_{1}) \rangle = \langle
L(u,p_{2},\bar{p}_{2}) \rangle. \label{eq:16}
\end{equation}

We obtain from this by comparing coefficients of the different
powers of $u$:
\begin{equation}
\delta_{1} = \delta_{2}, \qquad \beta_{1} = \beta_{2}, \qquad
\phi_{1} - \psi_{1} = \phi_{2} - \psi_{2}, \label{eq:17}
\end{equation}
\begin{equation}
\frac{\lambda_{1}\phi_{1}}{\psi_{1}}=\frac{\lambda_{2}\phi_{2}}{\psi_{2}}.
\label{eq:18}
\end{equation}

In addition to this, from (\ref{eq:16}) it follows that
$$
\det \; \langle L(u,p_{1},\bar{p}_{1}) \rangle = \det \; \langle
L(u,p_{2},\bar{p}_{2}) \rangle.
$$

In this equation the left-hand and the right-hand sides are
polynomials of the second powers of $u$. The roots of the left
polynomial:
$$
u_{1} = \lambda_{1}, \qquad \bar{u}_{1}=\mu_{1}.
$$
The roots of the right polynomial:
$$
u_{2} = \lambda_{2}, \qquad \bar{u}_{2}=\mu_{2}.
$$

It is clear that the roots of the left-hand and the right-hand
sides must coincide. We consider the following case:
\begin{equation}
\lambda_{1}=\mu_{2}, \qquad \lambda_{2}=\mu_{1}. \label{eq:19}
\end{equation}

Rewriting (\ref{eq:17})$-$(\ref{eq:19}) in terms of $a_{i}$,
$b_{i}$, $c_{i}$, $d_{i}$, we obtain
$$
\left\{
\begin{array}{l}
a_{1}^{N} \bar{a}_{1}^{N} = a_{2}^{N}\bar{a}_{2}^{N}, \qquad
b_{1}^{N} \bar{b}_{1}^{N} = b_{2}^{N}\bar{b}_{2}^{N},\\
\vspace{-2mm}\\
\displaystyle \frac{\bar{c}_{1}^{N}\bar{d}_{1}^{N}}
{\bar{a}_{1}^{N} \bar{b}_{1}^{N}} =
\frac{c_{2}^{N}d_{2}^{N}}{a_{2}^{N}b_{2}^{N}}, \qquad
\frac{\bar{c}_{2}^{N} \bar{d}_{2}^{N}}{\bar{a}_{2}^{N}
\bar{b}_{2}^{N}} =
\frac{c_{1}^{N}d_{1}^{N}}{a_{1}^{N}b_{1}^{N}},\\
\vspace{-2mm}\\
\displaystyle \frac{d_{1}^{N} \bar{d}_{1}^{N}}{a_{1}^{N}
\bar{a}_{1}^{N}} =
\frac{d_{2}^{N} \bar{d}_{2}^{N}}{a_{2}^{N} \bar{a}_{2}^{N}},\\
\vspace{-2mm}\\
d_{1}^{N} \bar{b}_{1}^{N} - a_{1}^{N} \bar{c}_{1}^{N} = d_{2}^{N}
\bar{b}_{2}^{N} - a_{2}^{N} \bar{c}_{2}^{N}.
\end{array}
\right.
$$

We now consider two representations of the $L$-operators algebra:
$L_{1}(u,p_{1},\bar{p}_{1})L_{2}(u,p_{2},\bar{p}_{2})$
and\\ 
$L_{1}(u,p_{3},\bar{p}_{3})L_{2}(u,p_{4},\bar{p}_{4})$. We want
to find the conditions under which these two representations are
equivalent.

We have:
\begin{equation}
\begin{array}{l}
\det \langle L_{1}(u,p_{1},\bar{p}_{1}) \rangle \cdot \det
\langle L_{2}(u,p_{2},\bar{p}_{2}) \rangle = \\
\vspace{-2mm}\\
\qquad \qquad \qquad = \det \langle L_{1}(u,p_{3},\bar{p}_{3})
\rangle \cdot \det \langle L_{2}(u,p_{4},\bar{p}_{4}) \rangle.
\end{array}
\label{eq:a3}
\end{equation}

Each determinant is a square polynomial with regard to $u$. It is
not difficult to find its roots. They are
$$
u = \lambda,\; \bar{u}= \mu.
$$

As the result, we see that the left-hand side of (\ref{eq:c1})
vanishes if $u_{1} = \lambda_{1}$, $\bar{u}_{1}=\mu_{1}$,
$u_{2}=\lambda_{2}$, $\bar{u}_{2}=\mu_{2}$, and the right-hand
side vanishes if $u_{3}=\lambda_{3}$, $\bar{u}_{3}=\mu_{3}$,
$u_{4}=\lambda_{4}$, $\bar{u}_{4}=\mu_{4}$.

It is clear that the left and the right roots coincide. We
consider the following case:
$$
\mu_{2} = \lambda_{3}, \qquad \mu_{1} = \mu_{3}, \qquad
\lambda_{2} = \lambda_{4}, \qquad \lambda_{1} = \mu_{4}.
$$

We have:
\begin{equation}
\langle L_{1}(u,p_{1},\bar{p}_{1}) \rangle \langle
L_{2}(u,p_{2},\bar{p}_{2}) \rangle = \langle
L_{1}(u,p_{3},\bar{p}_{3}) \rangle \langle
L_{2}(u,p_{4},\bar{p}_{4}) \rangle,
\label{eq:b3}
\end{equation}
$$
\mu_{1}, \; \lambda_{1}  \qquad \qquad \mu_{2}, \; \lambda_{2}
\qquad \qquad \mu_{1}, \; \mu_{2} \qquad \qquad \lambda_{1}, \;
\lambda_{2}
$$
that is two roots $\lambda_{1}$ and $\mu_{2}$ change trade.

Let $u=\lambda_{2}$. We act from the right on the vector
$\Psi_{1}$ (the right zero vector of the operator $\langle
L_{2}(\lambda_{2},p_{2},\bar{p}_{2}) \rangle$) by the both sides
of the equation (\ref{eq:b3}):
$$
\Psi_{1}= \left(
\begin{array}{l}
-\lambda_{2} \delta_{2}\\
\vspace{-2mm}\\
\, \; \psi_{2}
\end{array}
\right), \qquad \langle L_{2}(\lambda_{2},p_{2},\bar{p}_{2})
\rangle \Psi_{1} = 0.
$$
This vector is the right zero vector for $\langle
L_{2}(\lambda_{2},p_{4},\bar{p}_{4}) \rangle$ too. Consequently,
the following equation is valid:
$$
\frac{\delta_{2}}{\psi_{2}}=\frac{\delta_{4}}{\psi_{4}}.
$$

Let now $u=\mu_{1}$. We act from the left on the left zero vector
$\Psi_{2}$ of the operator $\langle L_{1}(\mu_{1}, p_{1},
\bar{p}_{1}) \rangle$ by the both sides of the equation:
$$
\Psi_{2}=\left(
\begin{array}{l}
\phi_{1}\\
\vspace{-2mm}\\
\mu_{1} \beta_{1}
\end{array}
\right), \qquad \Psi_{2} \langle L_{1}(\mu_{1},p_{1},\bar{p}_{1})
\rangle = 0.
$$

Since this vector is the left zero vector of the operator
$\langle L_{1}(\mu,p_{3},\bar{p}_{3}) \rangle$, we have
$$
\frac{\phi_{1}}{\beta_{1}}=\frac{\phi_{3}}{\beta_{3}}.
$$

It is clear that (\ref{eq:b3}) is valid if we insert between the
two factors in the right-hand side the unity $1=M M^{-1}$, where
$M$ is a two-dimensional matrix:
$$
M=\left[
\begin{array}{ll}
m_{1} & 0\\
\vspace{-2mm}\\
0 & m_{2}
\end{array}
\right].
$$

Using this gauge symmetry, one can set
$$
\beta_{3}=\beta_{1}, \qquad \delta_{4}=\delta_{2},
$$
from which it follows immediately that
$$
\phi_{3}=\phi_{1}, \qquad \psi_{4}=\psi_{2}.
$$

Multiplying the matrices in (\ref{eq:b3}) and comparing
coefficients of the powers of $u$, it is not difficult to see that
$$
\left\{
\begin{array}{l}
\beta_{4} = \beta_{2}, \qquad \delta_{3}=\delta_{1},\\
\vspace{-2mm}\\
\displaystyle \psi_{3} = \phi_{4} \frac{\mu_{2}
\psi_{1}}{\lambda_{1} \phi_{2}},\\
\vspace{-2mm}\\
\psi_{3} \beta_{2} - \delta_{1} \phi_{4} = \psi_{1} \beta_{2} -
\delta_{1} \phi_{2}.
\end{array}
\right.
$$

From the two last equations we obtain
\begin{equation}
\phi_{4} = \frac{\lambda_{1} \phi_{2}(\beta_{2}
\psi_{1}-\delta_{1}\phi_{2})}{\beta_{2}\mu_{2}\psi_{1}
-\lambda_{1}\delta_{1}\phi_{2}}, \qquad \psi_{3} =
\frac{\mu_{2}\psi_{1}(\beta_{2}\psi_{1}-\delta_{1}\phi_{2})}
{\beta_{2}\mu_{2}\psi_{1}-\lambda_{1}\delta_{1}\phi_{2}}.
\label{eq:ñ3}
\end{equation}

Collecting all obtained equations we have in addition to (\ref{eq:ñ3})
$$
\beta_{1}=\beta_{3}, \qquad \beta_{2}=\beta_{4}, \qquad
\delta_{1} = \delta_{3}, \qquad \delta_{2} = \delta_{4},
$$
$$
\phi_{1} = \phi_{3}, \qquad \psi_{2}=\psi_{4}, \qquad
\mu_{2}=\lambda_{3},
$$
$$
\mu=\mu_{3}, \qquad \lambda_{2}=\lambda_{4}, \qquad
\lambda_{1}=\mu_{4}.
$$

\section{The operator $F$.}
\label{D}

The matrix equation (\ref{eq:13}) can be written in the form of a
system of equations corresponding to the four matrix elements. We have:
$$
\begin{array}{lll}
L_{1}(p_{1},\bar{p}_{1}) L_{2}(p_{2},\bar{p}_{2}) &=& \left[
\begin{array}{ll}
A_{1} & B_{1}\\
\vspace{-2mm}\\
C_{1} & D_{1}
\end{array}
\right] \cdot \left[
\begin{array}{ll}
A_{2} & B_{2}\\
\vspace{-2mm}\\
C_{2} & D_{2}
\end{array}
\right] = \\
\vspace{-2mm}\\
&=& \left[
\begin{array}{ll}
A_{1}A_{2}+B_{1}C_{2} & A_{1}B_{2}+B_{1}D_{2}\\
\vspace{-2mm}\\
C_{1}A_{2}+D_{1}C_{2} & C_{1}B_{2}+D_{1}D_{2}
\end{array}
\right].
\end{array}
$$

As the result, we have the system:
$$
\left\{
\begin{array}{l}
F(X_{1}X_{2}^{-1}) [A_{1}A_{2}+B_{1}C_{2}] =
[A_{3}a_{4}+B_{3}C_{4}]F(X_{1}X_{2}^{-1}),\\
\vspace{-2mm}\\
F(X_{1}X_{2}^{-1}) [A_{1}B_{2}+B_{1}D_{2}] =
[A_{3}B_{4}+B_{3}D_{4}]F(X_{1}X_{2}^{-1}),\\
\vspace{-2mm}\\
F(X_{1}X_{2}^{-1}) [C_{1}A_{2}+D_{1}C_{2}] =
[C_{3}a_{4}+D_{3}C_{4}]F(X_{1}X_{2}^{-1}),\\
\vspace{-2mm}\\
F(X_{1}X_{2}^{-1}) [C_{1}B_{2}+D_{1}D_{2}] =
[C_{3}B_{4}+D_{3}D_{4}]F(X_{1}X_{2}^{-1}).
\end{array}
\right.
$$

We choose a basis $|k_{1},k_{2}\rangle$, $k_{1},k_{2}=0,...,N-1$
(mod $N$):
$$
\begin{array}{l}
X_{1}|k_{1},k_{2} \rangle = \omega^{k_{1}}|k_{1},k_{2}\rangle,
\qquad X_{2}|k_{1},k_{2}\rangle = \omega^{k_{2}}
|k_{1},k_{2}\rangle,\\
\vspace{-2mm}\\
Z_{1}|k_{1},k_{2}\rangle = |k_{1}-1,k_{2}\rangle, \qquad
Z_{2}|k_{1},k_{2}\rangle = |k_{1},k_{2}-1 \rangle.
\end{array}
$$

In this basis the matrix $F(X_{1}X_{2}^{-1})$ is diagonal. Let us
computate its nonzero matrix elements.

Substituting the expressions for $A_{i}$, $B_{i}$, $C_{i}$,
$D_{i}$, $i=1,2$, we have, for example, for the first equation:
$$
\begin{array}{l}
F(X_{1}X_{2}^{-1}) \left[
(c_{1}\bar{c}_{1}Z_{1}-b_{1}\bar{b}_{1}u)(c_{2}\bar{c}_{2}Z_{2}-b_{2}\bar{b}_{2}u)-
\right.\\
\vspace{-2mm}\\
\left. \qquad \qquad
-u(b_{1}\bar{d}_{1}-c_{1}\bar{a}_{1}Z_{1})X_{1}X_{2}^{-1}(d_{2}\bar{b}_{2}-a_{2}\bar{c}_{2}Z_{2})
\right]=\\
\vspace{-2mm}\\
\qquad =\left[
(c_{3}\bar{c}_{3}Z_{1}-b_{3}\bar{b}_{3}u)(c_{4}\bar{c}_{4}Z_{2}-b_{4}\bar{b}_{4}u)- \right.\\
\vspace{-2mm}\\
\qquad \qquad \left.
-u(b_{3}\bar{d}_{3}-c_{3}\bar{a}_{3}Z_{1})X_{1}X_{2}^{-1}(d_{4}\bar{b}_{4}-a_{4}\bar{c}_{4}Z_{2})
\right] F(X_{1}X_{2}^{-1}).
\end{array}
$$

Opening the parenthesis and acting by the left-hand and the
right-hand sides on the vector $|k_{1},k_{2}\rangle$ and comparing
coefficents of the linearly independent vectors and different
powers of $u$, we obtain:
$$
\left\{
\begin{array}{l}
c_{1}\bar{c}_{1}c_{2}\bar{c}_{2} = c_{3}\bar{c}_{3}c_{4}\bar{c}_{4},\\
\vspace{-2mm}\\
b_{1}\bar{b}_{1}b_{2}\bar{b}_{2} = b_{3}\bar{b}_{3}b_{4}\bar{b}_{4},\\
\vspace{-2mm}\\
b_{1}\bar{d}_{1}d_{2}\bar{b}_{2} = b_{3}\bar{d}_{3}d_{4}\bar{b}_{4},\\
\vspace{-2mm}\\
c_{1}\bar{a}_{1}a_{2}\bar{c}_{2} = c_{3}\bar{a}_{3}a_{4}\bar{c}_{4},\\
\vspace{-2mm}\\
\displaystyle F(\omega^{k+1}) =
\frac{b_{3}\bar{c}_{4}(\bar{d}_{3}a_{4}\omega^{k+1}-\bar{b}_{3}c_{4})}
{b_{1}\bar{c}_{2}(\bar{d}_{1}a_{2}\omega^{k+1}-\bar{b}_{1}c_{2})}
F(\omega^{k})
=\frac{c_{1}\bar{b}_{2}(\bar{a}_{1}d_{2}\omega^{k+1}-\bar{c}_{1}b_{2})}
{c_{3}\bar{b}_{4}(\bar{a}_{3}d_{4}\omega^{k+1}-\bar{c}_{3}b_{4})}F(\omega^{k}),
\end{array}
\right.
$$
where $F(\omega^{k})$, $k=0,...,N-1$, are the diagonal matrix
elements of the matrix $F$.

Using the obtained restrictions on $a_{i}$, $b_{i}$, $c_{i}$,
$d_{i}$, we can reduce this system to the smaller one:
$$
\left\{
\begin{array}{l}
\bar{c}_{1}c_{2}=\bar{c}_{3}c_{4},\\
\vspace{-2mm}\\
\bar{d}_{1}d_{2}=\bar{d}_{3}d_{4},\\
\vspace{-2mm}\\
\displaystyle
F(\omega^{k+1})=\frac{\bar{d}_{3}a_{4}\omega^{k+1}-\bar{b}_{3}c_{4}}
{\bar{d}_{1}a_{2}\omega^{k+1}-\bar{b}_{1}c_{2}} F(\omega^{k})
=\frac{\bar{a}_{1}d_{2}\omega^{k+1}-\bar{c}_{1}b_{2}}
{\bar{a}_{3}d_{4}\omega^{k+1}-\bar{c}_{3}b_{4}}F(\omega^{k})
\end{array}
\right.
$$

One can easily see that the second equation can be derived from
the first if we use the constraints
$$
\bar{c}_{3}\bar{d}_{3}=c_{2}d_{2}\frac{\bar{a}_{3}\bar{b}_{3}}{a_{2}b_{2}},
\qquad
c_{4}d_{4}=\bar{c}_{1}\bar{d}_{1}\frac{a_{2}b_{2}}{\bar{a}_{1}\bar{b}_{1}}.
$$

Since $F(\omega^{k})$ has a single meaning, we must to equal the
two fractions in terms of which $F(\omega^{k})$ is expressed. We
obtain:
$$
(\bar{d}_{3}a_{4}\omega^{k+1}-\bar{b}_{3}c_{4})(\bar{a}_{3}d_{4}\omega^{k+1}-\bar{c}_{3}b_{4})=
(\bar{d}_{1}a_{2}\omega^{k+1}-\bar{b}_{1}c_{2})(\bar{a}_{1}d_{2}\omega^{k+1}-\bar{c}_{1}b_{2}).
$$
Comparing coefficients of different powers of $\omega$, we obtain:
$$
\left\{
\begin{array}{l}
\bar{a}_{1}\bar{d}_{1}a_{2}d_{2}=\bar{a}_{3}\bar{d}_{3}a_{4}d_{4},\\
\vspace{-2mm}\\
\bar{b}_{1}\bar{c}_{1}b_{2}c_{2}=\bar{b}_{3}\bar{c}_{3}b_{4}c_{4},\\
\vspace{-2mm}\\
\displaystyle \bar{a}_{1}\bar{b}_{1}a_{2}b_{2}\left(
\frac{\bar{c}_{1}\bar{d}_{1}}{\bar{a}_{1}\bar{b}_{1}} +
\frac{c_{2}d_{2}}{a_{2}b_{2}}\right)=a_{4}b_{4}\bar{a}_{3}\bar{b}_{3}
\left( \frac{\bar{c}_{3}\bar{d}_{3}}{\bar{a}_{3}\bar{b}_{3}} +
\frac{c_{4}d_{4}}{a_{4}b_{4}} \right).
\end{array}
\right.
$$

If we take into account the restrictions on $a_{i}$, $b_{i}$,
$c_{i}$, $d_{i}$ again, then we have from this:
\begin{equation}
\bar{c}_{1}c_{2}=\bar{c}_{3}c_{4}, \label{eq:14}
\end{equation}
that is, the same that earlier.

We also have from the evident equation
$$
F(\omega^{N+k})=F(\omega^{k})
$$
that
\begin{equation}
\bar{c}_{1}^{N} b_{2}^{N} -
\bar{a}_{1}^{N}d_{2}^{N}=\bar{c}_{3}^{N}b_{4}^{N}-\bar{a}_{3}^{N}d_{4}^{N}.
\label{eq:15}
\end{equation}

We now prove that (\ref{eq:14}), (\ref{eq:15}) follow from
(\ref{eq:8})$-$(\ref{eq:12}). We have:
$$
\psi_{3}=a_{3}^{N}\bar{c}_{3}^{N}=\frac{c_{2}^{N}\bar{b}_{1}^{N}a_{1}^{N}}{b_{2}^{N}}
\frac{b_{2}^{N}
\bar{c}_{1}^{N}-\bar{a}_{1}^{N}d_{2}^{N}}{c_{2}^{N}\bar{b}_{1}^{N}-a_{2}^{N}\bar{d}_{1}^{N}},
$$
$$
\phi_{4}=d_{4}^{N}\bar{b}_{4}^{N}=\frac{\bar{d}_{1}^{N}a_{2}^{N}\bar{b}_{2}^{N}}{\bar{a}_{1}^{N}}
\frac{b_{2}^{N}
\bar{c}_{1}^{N}-\bar{a}_{1}^{N}d_{2}^{N}}{c_{2}^{N}\bar{b}_{1}^{N}-a_{2}^{N}\bar{d}_{1}^{N}}.
$$
Expressing from this $\bar{c}_{3}^{N}$ and $d_{4}^{N}$ and
substituting them into (\ref {eq:15}), we obtain the identity.
Apart from, we have:
$$
\frac{\bar{c}_{3}}{d_{4}} =
\frac{c_{2}}{\bar{d}_{1}}\frac{\bar{a}_{1}\bar{b}_{1}}{a_{2}b_{2}}.
$$
Multiplying the last equation by
$$
c_{4}d_{4} =
\bar{c}_{1}\bar{d}_{1}\frac{a_{2}b_{2}}{\bar{a}_{1}\bar{b}_{1}},
$$
we obtain (\ref{eq:14}).

Thus,
$$
F(\omega^{k+1})=\sqrt[N]{\frac{\bar{d}_{1}^{N}a_{2}^{N}-\bar{b}_{1}^{N}c_{2}^{N}}
{\bar{a}_{1}^{N}d_{2}^{N}-\bar{c}_{1}^{N}b_{2}^{N}}}
\frac{\bar{c}_{1}b_{2}-\bar{a}_{1}d_{2}\omega^{k+1}}{\bar{b}_{1}c_{2}-\bar{d}_{1}a_{2}\omega^{k+1}}
F(\omega^{k}).
$$

\section{The star-triangle equation.}
\label{A}

Here we prove the star-triangle equation,
\begin{equation}
G(Z_{1}) F(X_{1}X_{2}^{-1}) G(Z_{1}) = \mu F(X_{1}X_{2}^{-1})
G(Z_{1}) F(X_{1}X_{2}^{-1}), \label{eq:a1}
\end{equation}
and find $\mu$.

So, we have:
$$
(p,q)(r,s) \stackrel{F}{\longrightarrow} (p,r')(q',s)
\stackrel{G}{\longrightarrow} (r'',p')(q',s)
$$
$$
\downarrow G \qquad \qquad \qquad \qquad \qquad \qquad F
\downarrow
$$
$$
(\tilde{q},\tilde{p})(r,s) \stackrel{F}{\longrightarrow}
(\tilde{q},\tilde{r})(p'',s) \stackrel{G}{\longrightarrow}
(r'',q'')(p'',s)
$$

Now if $r''$, $q''$, $\tilde{p'}$, obtained in the two different
ways, coincide exactly within the gauge trasnformations, then
the equation (\ref{eq:a1}) is actually valid.

We recall the definition of the fuctions
$\Lambda(p_{1},p_{2})$, $\Omega(p_{1},p_{2})$:
\begin{equation}
\Lambda(p_{1},p_{2})=\sqrt[N]{\frac{b_{1}^{N}d_{2}^{N} -
c_{1}^{N}a_{2}^{N}}{d_{1}^{N}b_{2}^{N} - a_{1}^{N}c_{2}^{N}}}.
\label{eq:c1}
\end{equation}
\begin{equation}
\Omega(p_{1},p_{2}) = \sqrt[N]{\frac{c_{1}^{N}b_{2}^{N} -
a_{1}^{N}d_{2}^{N}}{b_{1}^{N}c_{2}^{N} - d_{1}^{N}a_{2}^{N}}},
\label{eq:b1}
\end{equation}

Let us make all computations for the first chain:
\begin{enumerate}
\item[1.]$(p,q)(r,s) \stackrel{F}{\longrightarrow}
(p,r')(q',s)$.

We have:
$$
\begin{array}{ll}
a'_{r} = a_{r}, & a'_{q} = a_{q},\\
\vspace{-2mm}\\
b'_{r} = b_{r}, & b'_{q} = b_{q},\\
\vspace{-2mm}\\
c'_{r} = \Omega(q,r) c_{r}, & c'_{q} = \Omega(q,r)^{-1} c_{q},\\
\vspace{-2mm}\\
d'_{r} = \Omega(q,r)^{-1} d_{r}, & d'_{q} = \Omega(q,r)d_{q}.
\end{array}
$$

\item[2.]$(p,r')(q',s)
\stackrel{G}{\longrightarrow} (r'',p')(q',s)$.

We have:
$$
\begin{array}{ll}
a''_{r} = a'_{r}, & a'_{p} = a_{p},\\
\vspace{-2mm}\\
b''_{r} = b'_{r}, & b'_{p} = b_{p},\\
\vspace{-2mm}\\
c''_{r} = \Lambda(p,r') c'_{r}, & c'_{p} = \Lambda(p,r')^{-1} c_{p},\\
\vspace{-2mm}\\
d''_{r} = \Lambda(p,r')^{-1} d'_{r}, & d'_{p} =
\Lambda(p,r')d_{p}.
\end{array}
$$

\item[3.]$(r'',p')(q',s) \stackrel{F}{\longrightarrow}
(r'',q'')(p'',s)$.

We have:
$$
\begin{array}{ll}
a''_{q} = a'_{q}, & a''_{p} = a'_{p},\\
\vspace{-2mm}\\
b''_{q} = b'_{q}, & b''_{p} = b'_{p},\\
\vspace{-2mm}\\
c''_{q} = \Omega(p',q') c'_{q}, & c''_{p} = \Omega(p',q')^{-1} c'_{p},\\
\vspace{-2mm}\\
d''_{q} = \Omega(p',q')^{-1} d'_{q}, & d''_{p} =
\Omega(p',q')d'_{p}.
\end{array}
$$
\end{enumerate}

Now we make all computations for the second chain:
\begin{enumerate}
\item[1.]$(p,q)(r,s) \stackrel{G}{\longrightarrow}
(\tilde{q},\tilde{p})(r,s)$.

We have:
$$
\begin{array}{ll}
\tilde{a}_{q} = a_{q}, & \tilde{a}_{p} = a_{p},\\
\vspace{-2mm}\\
\tilde{b}_{q} = b_{q}, & \tilde{b}_{p} = b_{p},\\
\vspace{-2mm}\\
\tilde{c}_{q} = \Lambda(p,q) c_{q}, & \tilde{c}_{p} = \Lambda(p,q)^{-1} c_{p},\\
\vspace{-2mm}\\
\tilde{d}_{q} = \Lambda(p,q)^{-1} d_{q}, & \tilde{d}_{p} =
\Lambda(p,q)d_{p}.
\end{array}
$$

\item[2.]$(\tilde{q},\tilde{p})(r,s)
\stackrel{F}{\longrightarrow} (\tilde{q},\tilde{r})(p'',s)$.

We have:
$$
\begin{array}{ll}
\tilde{a}_r = a_{r}, & a''_{p} = \tilde{a}_{p},\\
\vspace{-2mm}\\
\tilde{b}_r = b_{r}, & b''_{p} = \tilde{b}_{p},\\
\vspace{-2mm}\\
\tilde{c}_r = \Omega(\tilde{p},r) c_{r}, & c''_{p} = \Omega(\tilde{p},r)^{-1} \tilde{c}_{p},\\
\vspace{-2mm}\\
\tilde{d}_r = \Omega(\tilde{p},r)^{-1} d_{r}, & d''_{p} =
\Omega(\tilde{p},r)\tilde{d}_{p}.
\end{array}
$$

\item[3.]$(\tilde{q},\tilde{r})(p'',s) \stackrel{G}{\longrightarrow}
(r'',q'')(p'',s)$.

We have:
$$
\begin{array}{ll}
a''_{r} = \tilde{a}_r, & a''_{q} = \tilde{a}_{q},\\
\vspace{-2mm}\\
b''_{r} = \tilde{b}_r, & b''_{q} = \tilde{b}_{q},\\
\vspace{-2mm}\\
c''_{r} = \Lambda(\tilde{q},\tilde{r}) \tilde{c}_r, & c''_{q} =
\Lambda(\tilde{q},\tilde{r})^{-1} \tilde{c}_{q},\\
\vspace{-2mm}\\
d''_{r} = \Lambda(\tilde{q},\tilde{r})^{-1} \tilde{d}_r, & d''_{q}
= \Lambda(\tilde{q},\tilde{r})\tilde{d}_{q}.
\end{array}
$$
\end{enumerate}

It remains to verify that the obtained $L$-operators actually
coincide. Comparing the parameters $r''$, $q''$, $p''$, obtained
in the two different ways, we conclude that the equation
(\ref{eq:a}) satisfies if
\begin{equation}
\begin{array}{l}
\Lambda(p,r') \Omega(p',q') = \Lambda(p,q)
\Omega(\tilde{p},r),\\
\vspace{-2mm}\\
\Omega(q,r) \Lambda(p,r') = \Omega(\tilde{p},r)
\Lambda(\tilde{q},\tilde{r}).
\end{array}
\label{eq:d1}
\end{equation}

It is not difficult to show that (\ref{eq:d1}) satisfies
identically. Thus, we prove (\ref{eq:a1}) for some unknown 
$\mu$. We now find $\mu^{N}$.

It is clear that
\begin{equation}
\mu^{N} = \frac{\det \; G(p,q) \cdot \det \; F(\tilde{p},r) \cdot
\det \; G(\tilde{q},\tilde{r})}{\det \; F(q,r) \cdot \det \;
G(p,r') \cdot \det \; F(p',q')}. \label{eq:e1}
\end{equation}

One can derive the determinants $G(p_{1},p_{2})$ and $F(p_{1},p_{2})$ 
without difficult. Each of these matrices can be reduced to the 
diagonal form (not at the same time), with in each case
the diagonal matrix elements being given by the following
formulae:
$$
\begin{array}{l}
\displaystyle \frac{G(p_{1},p_{2}; \omega^{k})}{G(p_{1},p_{2}; 1)}
= \left( \frac{b_{1}^{N} d_{2}^{N} - c_{1}^{N}
a_{2}^{N}}{d_{1}^{N} b_{2}^{N} - a_{1}^{N} c_{2}^{N}}
\right)^{k/N} \prod_{j=1}^{k} \frac{d_{1}b_{2} -
a_{1}c_{2}\omega^{j}}{b_{1}d_{2} - c_{1}a_{2}\omega^{j}},\\
\vspace{-2mm}\\
\displaystyle \frac{F(p_{1},p_{2}; \omega^{k})}{F(p_{1},p_{2}; 1)}
= \left( \frac{b_{1}^{N}c_{2}^{N} - d_{1}^{N}
a_{2}^{N}}{c_{1}^{N} b_{2}^{N} - a_{1}^{N} d_{2}^{N}}
\right)^{k/N} \prod_{j=1}^{k} \frac{c_{1} b_{2} - a_{1}
d_{2}\omega^{j}}{b_{1} c_{2} - d_{1} a_{2} \omega^{j}}.
\end{array}
$$

We set
$$
G(p_{1},p_{2}; 1) = F(p_{1},p_{2}; 1) = 1.
$$

Then
$$
\begin{array}{l}
\displaystyle \det \; G(p_{1},p_{2}) = \left( \frac{b_{1}^{N}
d_{2}^{N} - c_{1}^{N} a_{2}^{N}}{d_{1}^{N} b_{2}^{N} - a_{1}^{N}
c_{2}^{N}} \right)^{\frac{N-1}{2}} \prod_{k=1}^{N-1}
\prod_{j=1}^{k} \frac{d_{1}b_{2} -
a_{1}c_{2}\omega^{j}}{b_{1}d_{2} -
c_{1}a_{2}\omega^{j}},\\
\vspace{-2mm}\\
\displaystyle \det \; F(p_{1},p_{2})= \left(
\frac{b_{1}^{N}c_{2}^{N} - d_{1}^{N} a_{2}^{N}}{c_{1}^{N}
b_{2}^{N} - a_{1}^{N} d_{2}^{N}} \right)^{\frac{N-1}{2}}
\prod_{k=1}^{N-1} \prod_{j=1}^{k} \frac{c_{1} b_{2} - a_{1}
d_{2}\omega^{j}}{b_{1} c_{2} - d_{1} a_{2} \omega^{j}}.
\end{array}
$$

\section{The relations in the algebra of the $Q$-operators.}
\label{AlQ}

The relations in the $Q$-operators algebra follow from the
properties of the cyclic representations of the $L$-operators 
algebra. We now prove this.

The relations (\ref{eq:Sootn0a}), (\ref{eq:Sootn0b}),
(\ref{eq:Sootn3a}), (\ref{eq:Sootn3b}) become evident if we
recall that
$$
Q(u) = {\rm tr}_{0} L_{1 0}(u)L_{2 0}(u)...L_{n 0}(u),
$$
and the following symmetries are the case:
$$
L(\lambda p_{1}, \bar{p}_{1}) = \lambda L(p_{1}, \bar{p}_{1}),
$$
$$
L(p_{1}, \mu \bar{p}_{1}) = \mu L(p_{1},\bar{p}_{1}),
$$
$$
L(a_{1},b_{1},c_{1},d_{1},\bar{a}_{1}, \bar{b}_{1}, \bar{c}_{1},
\bar{d}_{1}) = L(\lambda a_{1}, b_{1}, \lambda c_{1}, d_{1},
\lambda^{-1} \bar{a}_{1}, \bar{b}_{1}, \lambda^{-1} c_{q},
\bar{d}_{1}),
$$
$$
L(a_{1},b_{1},c_{1},d_{1},\bar{a}_{1}, \bar{b}_{1}, \bar{c}_{1},
\bar{d}_{1}) = L(a_{1}, \mu b_{1}, c_{1}, \mu d_{1}, \bar{a}_{1},
\mu^{-1} \bar{b}_{1}, \bar{c}_{1}, \mu^{-1} \bar{d}_{1}),
$$
where $\lambda$, $\mu$ are arbitrary numbers.

The relation (\ref{eq:1}) follows from another symmetry:
$$
L_{1}(p_{1},\bar{p}_{1})L_{2}(p_{2},s) = L_{1}(p_{1},\bar{p}_{1})
M M^{-1} L_{2}(p_{2},s),
$$
or, in more details,
$$
\begin{array}{l}
L_{1}(a_{1},b_{1},c_{1},d_{1},\bar{a}_{1},
\bar{b}_{1},\bar{c}_{1},\bar{d}_{1})
L_{2}(a_{2},b_{2},c_{2},d_{2},\bar{a}_{2},
\bar{b}_{2},\bar{c}_{2},\bar{d}_{2}) = \\
\vspace{-2mm}\\
\qquad \qquad \qquad = L_{1}(a_{1},b_{1},c_{1},d_{1},\beta
\bar{a}_{1},
\alpha \bar{b}_{1},\alpha \bar{c}_{1},\beta \bar{d}_{1})\\
\vspace{-2mm}\\
\qquad \qquad \qquad \qquad \times
L_{2}(\beta^{-1}a_{2},\alpha^{-1}b_{2},\alpha^{-1}c_{2},\beta^{-1}
d_{2},\bar{a}_{2},\bar{b}_{2},\bar{c}_{2},\bar{d}_{2}),
\end{array}
$$
where $\alpha$, $\beta$ are arbitrary numbers. Here
$$
M = \left(
\begin{array}{ll}
\alpha & 0\\
\vspace{-2mm}\\
0 & \beta
\end{array}
\right).
$$

The relation (\ref{eq:2}) is obtained from the equation
$$
G L(a_{1}, b_{1}, c_{1}, d_{1}, \bar{a}_{1}, \bar{b}_{1},
\bar{c}_{1}, \bar{d}_{1})G^{-1} = L(a_{1}, b_{1}, \Lambda c_{1},
\Lambda^{-1} d_{1}, \bar{a}_{1}, \bar{b}_{1}, \Lambda^{-1}
\bar{c}_{1}, \Lambda \bar{d}_{1}),
$$
where
$$
\Lambda = \Lambda(p_{1},\bar{p}_{1}) = \sqrt[N]{\frac{b_{1}^{N}
\bar{d}_{1}^{N} - c_{1}^{N} \bar{a}_{1}^{N}}{d_{1}^{N}
\bar{a}_{1}^{N} - a_{1}^{N} \bar{c}_{1}^{N}}},
$$
and the relation (\ref{eq:4}) is obtained from
$$
\begin{array}{l}
F L_{1}(a_{1},b_{1},c_{1},d_{1},\bar{a}_{1},
\bar{b}_{1},\bar{c}_{1},\bar{d}_{1})
L_{2}(a_{2},b_{2},c_{2},d_{2},\bar{a}_{2},
\bar{b}_{2},\bar{c}_{2},\bar{d}_{2}) F^{-1} =\\
\vspace{-2mm}\\
\qquad \qquad = L_{1}(a_{1},b_{1},c_{1},d_{1}, a_{2}, b_{2}, \Omega
c_{2},\Omega^{-1} d_{2})\\
\vspace{-2mm}\\ \qquad \qquad \qquad \qquad \times
L_{2}(\bar{a}_{1},\bar{b}_{1}, \Omega^{-1}\bar{c}_{1},\Omega
\bar{d}_{1}, \bar{a}_{2},\bar{b}_{2},\bar{c}_{2},\bar{d}_{2}),
\end{array}
$$
where
$$
\Omega = \Omega(\bar{p}_{1},p_{2}) =
\sqrt[N]{\frac{\bar{c}_{1}^{N} b_{2}^{N} - \bar{a}_{1}^{N}
d_{2}^{N}}{\bar{b}_{1}^{N} c_{2}^{N} - \bar{d}_{1}^{N}
a_{2}^{N}}}.
$$

\end{document}